\documentclass[10pt]{article}
\usepackage{epsfig}
\usepackage{amsfonts}
\usepackage{graphicx}
\usepackage{epsfig} 

\date{\today} 

\newcommand{\bm}{\begin{multiline}}
\newcommand{\beq}{\begin{equation}}
\newcommand{\eeq}{\end{equation}}
\newcommand{\beqs}{\begin{eqnarray}}
\newcommand{\eeqs}{\end{eqnarray}}

\newcommand{\tr}{\mbox{\rm tr}}
\newcommand{\ra}{\rightarrow}

\begin{document}

\title{Cosmological monopoles and non-abelian black holes}
\author{{\large Yves Brihaye }$^{\dagger}$,
{\large Betti Hartmann}$^{\ddagger}$,
{\large Eugen Radu}$^{\dagger \star}$ 
and {\large Cristian Stelea}$^{\ddagger \star}$ 
\\ \\
$^{\dagger}${\small Facult\'e des Sciences, Universit\'e de Mons-Hainaut,
B-7000 Mons, Belgium } \\
$^{\ddagger}${\small School of Engineering and Science, International University Bremen,
28725 Bremen, Germany} \\
$^{\dagger \star}${\small Department of  Mathematical Physics,
National University of Ireland Maynooth, Ireland}\\
$^{\ddagger \star}${\small Department of Physics, University of Waterloo,
Ontario N2L 3G1, Canada}
}

\date{\today}

\maketitle
 
\begin{abstract} 
We discuss magnetic monopole solutions 
of the Einstein-Yang-Mills-Higgs equations 
with a positive cosmological constant.
These configurations 
approach asymptotically the de Sitter spacetime background
and exist only for a nonzero Higgs potential.
We find that the total mass of the solutions within the cosmological
horizon is finite.
However, 
their mass evaluated by using the surface counterterm method
outside the cosmological
horizon at early/late time infinity 
generically diverges.
Magnetic monopole solutions  with finite mass  and noninteger charge
exist however
in a truncation of the theory with a vanishing Higgs field.
Both solutions with a regular origin  and  
cosmological black holes are studied, special attention being paid to
the
computation of the global charges.

\end{abstract}

\section{Introduction}
As found some time ago by 't Hooft and Polyakov, 
certain gauge theories  admit classical, particle-like
solutions with quantised charge and finite energy \cite{'tHooft:1974qc}.
The magnetic monopoles
inevitably arise in grand unification theories   whenever the spontaneous symmetry
breakdown of this theory generates a U(1) subgroup 
and are stabilised by a quantum number of topological
origin, corresponding to their magnetic charge.
 Since the Standard Model of particle physics contains a U(1) subgroup, monopoles
with masses of the order $10^{17}$ GeV are predicted which leads to the
so-called ``monopole problem''. The accepted solution to this problem
is the theory of inflation. 

When coupling the Yang-Mills-Higgs (YMH) system to gravity, a branch of 
globally regular monopoles
emerges smoothly from the corresponding flat space solutions.
The  nonabelian black hole solutions emerge from the globally regular configurations,
when a finite regular event horizon radius is imposed \cite{Brihaye:1999nn, Breitenlohner:1991aa}.
These solutions cease to exist beyond some maximal value of the coupling constant 
$\alpha$, which is proportional to the ratio of the vector meson mass and Planck mass.

It has been speculated that such configurations might have played
an important role in the early stages of the evolution of the
Universe. Also, various analyses indicate that the monopole 
solutions are important in quantum theories.
However, most of the investigations in the literature 
correspond to an asymptotically flat (AF) spacetime. 
Because of the physical importance of these objects, it is worthwhile to study
generalisations in a different cosmological background.

As discussed in
\cite{Lugo:1999fm},  the properties of magnetic monopole solutions 
in Anti-de Sitter (AdS) spacetime are rather similar to their
asymptotically flat counterparts.
Nontrivial solutions exist for any value of
the cosmological constant $\Lambda<0$;
as a new feature, one finds a complicated power decay of the fields
at infinity and a decrease of the
maximally allowed vacuum  expectation value of the Higgs field.

For a positive cosmological constant,
the natural ground state of the theory corresponds to
de Sitter (dS) spacetime.
Solutions of Einstein equations with this type
of asymptotics enjoyed recently a huge interest in theoretical physics for a
variety of reasons.
First at all, the observational evidence accumulated in the last
years (see, $e.g.$, ref.~\cite{data}), seems to favour the idea that the
physical universe has an accelerated expansion. The most common explanation
is that the expansion is driven by a small positive vacuum energy
($i.e.$ a cosmological constant $\Lambda>0$), implying
the spacetime is asymptotically dS. 
Furthermore, dS spacetime plays a central
role in the theory of inflation (the very rapid accelerated expansion in the
early universe), which is supposed to solve the cosmological flatness and
horizon problems.

Another motivation for studying solutions with this type of asymptotics is 
connected with the proposed holographic duality
between quantum gravity in dS spacetime and a 
conformal field theory (CFT) on the boundary of dS spacetime \cite{Strominger:2001pn}.
The results in the literature suggest that the conjectured dS/CFT correspondence
has a number of similarities with the AdS/CFT correspondence, although
many details and interpretations remain to be clarified
(see \cite{Klemm:2004mb} for a recent review and a large set 
of references on this problems).

In view of these developments, an examination of 
the classical solutions of gravitating 
fields in asymptotically  dS spacetimes seems appropriate,
the physical relevant case of a spontaneously  broken
nonabelian gauge theory being particularly interesting.

Considering a static coordinate system 
(which generalises for $\Lambda>0$ 
the usual Schwarzschild coordinates),
these solutions will present a cosmological horizon for a finite value
of the radial coordinate, where all curvature invariants stay finite.
Similar to the  well-known (electro)-vacuum solutions, 
 there is no global
timelike Killing vector, as the norm of the Killing vector $\partial/\partial t$
 changes sign as
it crosses the cosmological horizon. 
Inside the horizon, the Killing vector is timelike,
and this can be used to calculate the conserved charges and
action/entropy inside the
cosmological horizon \cite{Gibbons:1977mu}. 
Outside  the cosmological
horizon, however, with the change of $\partial/\partial t$ 
to a spacelike Killing vector, the
physical meaning of the conserved quantities  is  less clear.

Recently a method for computing conserved charges (and associated boundary
stress tensors) of  asymptotically dS spacetimes from data at early or late time infinity
was proposed \cite{Balasubramanian:2001nb}, 
in analogy with the prescription used in
 asymptotically  AdS spacetimes \cite{Balasubramanian:1999re}.  
This approach 
uses counterterms on spatial boundaries at early and late
times, yielding a finite  action for asymptotically dS spacetimes.
The boundary stress tensor on the
spacetime boundary can also be calculated, and a conserved charge (spacelike, due
to its association with the Killing vector $\partial /\partial t$) -- now
interpreted as the mass of the solutions, can be defined.
 
The main goal of this paper is to present a study  of the basic 
properties of the spherically symmetric
monopole solutions in Einstein-Yang-Mills-Higgs (EYMH) theory
with a positive cosmological constant.
(Solutions of the Abelian-Higgs model with $\Lambda>0$ have been studied in \cite{Ghezelbash:2002cc}.) 
Although an analytic  solution  to the  EYMH 
equations appears to be intractable,
we present both  analytical and numerical  arguments for the
existence of both solutions with a regular origin and cosmological black-hole,  
approaching asymptotically the dS background.
As found in \cite{Brihaye:2005ft}, 
the  features of the dS solutions of a spontaneously  broken
nonabelian gauge theory  
are rather different as compared to
the asymptotically flat or AdS counterparts.
The most interesting feature is that the mass of dS magnetic monopoles
 evaluated at timelike infinity according to the
 counterterm prescription generically diverges,
 although the total mass within the cosmological
horizon of these configurations is finite.
Also, no solutions exist in the absence of a Higgs potential.

However, in the last years it became clear that 
a nonasymptotically flat metric background 
may allow for nonabelian magnetic monopole solutions
even in the absence of a Higgs field.
For example, a charge-one monopole  solution has been found 
by Chamseddine and Volkov  \cite{Chamseddine:1997nm}
in the context of the ${\cal{N}}=4$ $D=4$  Freedman-Schwarz  gauged
supergravity \cite{Freedman:1978ra}.
This is one of the few analytically known 
configurations involving both non-abelian gauge fields and gravity,
the expression of the magnetic potential coinciding, in a suitable coordinate system,
 to that
of the Prasad-Sommerfield monopole \cite{Prasad:1975kr}.
A solution with many similar properties exists also \cite{Chamseddine:2001hk}
in a version of  ${\cal{N}}=4$ $d=5$ Romans' gauged 
supergravity model \cite{Romans:1985ps}
with a Liouville  dilaton potential.

As found in  \cite{Winstanley:1998sn, Bjoraker:2000qd},
monopole-type solutions exist 
even in a simple Einstein-Yang-Mills (EYM) theory with a
negative cosmological constant.
These asymptotically AdS nonabelian solutions 
have finite mass and an arbitrary value of the gauge potential at infinity,
$i.e.$ an arbitrary value of the magnetic charge.
There are also finite energy solutions for 
several intervals of the shooting parameter 
(the value of gauge function at the origin or at the event horizon), rather than discrete values  
and stable monopole solutions in which the gauge field has no zeros.

It is natural to conjecture the existence of similar
configurations for $\Lambda>0$.
Solutions with a regular origin of the  EYM-SU(2) system with positive
cosmological constant 
have been considered by several authors
(see \cite{Volkov:1996qj}-\cite{Torii:1995wv}, 
and also the systematic approach
in \cite{Breitenlohner:2004fp}).  
Similar to the AdS case, the asymptotic
value of the gauge potential for solutions with $\Lambda>0$
is not fixed,  implying the existence of a nonvanishing magnetic charge.
However, all asymptotically dS configurations are unstable.
In this paper we propose to reconsider the properties of these  EYM-$\Lambda$
solutions, viewing them as magnetic monopoles
in a model without a Higgs field.
Both solutions with a regular origin and cosmological black hole
configurations are discussed.

The mass and boundary stress-tensor as well as the thermodynamic quantities
of both EYMH and EYM solutions discussed in this paper are computed  by  
using the counterterm formalism proposed in \cite{Balasubramanian:2001nb}.
We also comment on the implications of these solutions for the conjectured
dS/CFT correspondence. 

 The paper is structured as follows:  in the next Section
we present the general framework and analyse the field equations and
boundary conditions. In Section 3 we present our numerical results.
In Section 4 we  compute  the solutions' 
mass and action, while in 
Section 5 we  comment  on the general picture in a different coordinate
system.
We conclude with Section 6, where our results are summarised.
The Appendix  generalises the construction of \cite{Gibbons:1993xt} for a positive
 effective cosmological constant.
We show that the BPS monopole of four-dimensional YMH theory 
continue to be exact solutions even after 
the inclusion of gravitational, electromagnetic and  dilatonic interactions, 
provided that certain non-minimal interactions are included.

\section{General framework}
\subsection{Action principle}
The action for a  gravitating non-Abelian $SU(2)$ gauge field
coupled to a triplet Higgs field is
\begin{eqnarray}
\label{lag0} 
I_{B}=\int_{\mathcal{M}}  d^4x \sqrt{-g} \left( \frac{1}{16\pi G}
(R-2 \Lambda) +\mathcal{L}_{M}\right) 
-\frac{1}{8\pi G}\int_{\partial\mathcal{M}} d^{3}x\sqrt{-h}K,
\end{eqnarray}
with  Newton's constant $G$ and 
\begin{eqnarray}
\label{lag} 
\mathcal{L}_{M}=-
\frac{1}{2} {\rm Tr} (F_{\mu\nu} F^{\mu\nu}) -\frac{1}{4}{\rm
Tr}(D_\mu \Phi D^\mu \Phi)-V(\Phi),
\end{eqnarray}
where $V(\Phi)=\frac{1}{4}\lambda{\rm Tr}(\Phi^2 - \eta^2)^2$ is the usual Higgs potential,
with $\eta$ the vacuum expectation value (vev).
The gauge field strength tensor is given by
\begin{eqnarray}
F_{\mu\nu}=\partial_{\mu}A_{\nu}-\partial_{\nu}A_{\mu}
+ie[A_{\mu},A_{\nu}],
\end{eqnarray}
with
$D_{\mu}=\partial_{\mu}+ie[A_{\mu},\ ]$ being the covariant derivative,
$A_\mu$ the gauge potential 
and $e$  the Yang-Mills coupling constant.

The last term in  (\ref{lag0})  is the Hawking-Gibbons surface term \cite{Gibbons:1976ue}, 
where $K$ is the trace 
of the extrinsic curvature for the boundary $\partial\mathcal{M}$ and $h$ is the induced 
metric of the boundary. Of course, this term does not affect the equations of motion 
but is important when 
computing the mass and action of solutions.

Apart from the cosmological constant, the theory
contains three mass scales: the Planck mass $M_{Pl}=1/\sqrt{G}$, the
gauge boson mass $M_W=ev$ and the mass $M_H=\sqrt{\lambda}\eta$ of
the Higgs field.

 Varying the action
(\ref{lag0}) with respect to $g^{\mu\nu}$, $A_{\mu}$ and $\Phi$ we
have the field equations
\begin{eqnarray}
\nonumber
\label{einstein-eqs} 
R_{\mu\nu}-\frac{1}{2}g_{\mu\nu}R +\Lambda g_{\mu\nu} &=& 8\pi
G~(T_{\mu\nu}^{(H)}+T_{\mu\nu}^{(YM)}),
\\
\label{field-eqs}
\frac{1}{\sqrt{-g}}D_{\mu}(\sqrt{-g} F^{\mu \nu}) &=&\frac{1}{4}
ie[\Phi,D^{\nu}\Phi],
\\
\nonumber
\label{Heqs} \frac{1}{\sqrt{-g}}D_{\mu}(\sqrt{-g}D^{\mu} \Phi)+\lambda (\Phi^2 -\eta^2) \Phi  &=&0,
\end{eqnarray}
where the stress-energy tensor is
\begin{eqnarray}
T_{\mu\nu}^{(H)}&=&{\rm Tr}(\frac{1}{2}D_\mu \Phi D_\nu \Phi 
    -\frac{1}{4} g_{\mu\nu} D_\alpha \Phi D^\alpha \Phi)   - g_{\mu\nu}V(\Phi),
\\
T_{\mu\nu}^{(YM)}&=&
    2{\rm Tr} 
    ( F_{\mu\alpha} F_{\nu\beta} g^{\alpha\beta}
   -\frac{1}{4} g_{\mu\nu} F_{\alpha\beta} F^{\alpha\beta})
\ . 
\end{eqnarray}
The EYM model is found in the limit of a vanishing Higgs field $\Phi=0$, $V(\Phi)=0$.

\subsection{The ansatz }%
For $\Lambda \leq 0$, the above field equations
present a large variety of solutions, 
including configurations with  axial  symmetry only 
\cite{Hartmann:2001ic}, \cite{Radu:2004ys}.
However, here we will restrict to  the spherically symmetric case.

When discussing physics in dS spacetime, 
there are many coordinate systems to choose among. 
In this work will consider 
mainly the static system. The advantage of these coordinates is their obvious simplicity and the time
independence. This coordinate  system  is also computationally more convenient, 
since we will deal with ordinary differential equations.
The disadvantage is that the expansion of the universe is not manifest.
(In Section 5 we will discuss our results in a cosmological coordinate system.
Both can be used to approach the timelike infinity and will lead to different 
form of the boundary metric.)

Therefore we consider a line element given by
\begin{equation} 
\label{metric}
ds^{2}=\frac{dr^{2}}{N(r)}+r^2(d\theta^{2}+\sin^{2}\theta d\varphi^{2})-
\sigma^2(r)N(r)dt^{2}
\end{equation}
where a convenient parametrisation of the 
metric function $N(r)$ is
\begin{equation} 
N(r)=1-\frac{2m(r)}{r}- \frac{\Lambda r^2}{3}~.
\end{equation}
The function $m(r)$ is usually interpreted
as the total  mass-energy within the radius $r$. 
For $m(r)=0$, the empty dS space
written in a static coordinate system with a
cosmological horizon at $r=r_c=\sqrt{3/\Lambda}$ is recovered. 

For the SU(2) Yang-Mills field we use the minimal spherically symmetric (purely
magnetic) ansatz employed also in previous studies on asymptotically flat (or AdS)
solutions, in terms of a single magnetic potential $w(r)$
\begin{eqnarray}
\label{A} 
A=\frac{1}{2e} \Big \{  
 \omega(r) \tau_1   d \theta
 +\Big(\cos \theta  \tau_3+  \omega(r)
\tau_2   \sin \theta  \Big)d \varphi \Big\},
\end{eqnarray}
while the  Higgs field is given by the usual form
\begin{eqnarray}
\label{higgs-form}
\Phi=\phi~\tau_3,
\end{eqnarray} 
where $\tau_i$ are the Pauli matrices.
\subsection{The equations of motion}%
For a nonvanishing asymptotic magnitude of the Higgs field
$\eta$, dimensionless quantities are obtained by considering 
the rescaling\footnote{It can be proven that
a vanishing asymptotic magnitude of the Higgs field $\eta$
 implies  $\Phi\equiv 0$.}  
\begin{eqnarray}
\label{rescaling}
 r\to r/(\eta e),~~~m\to m/(\eta e),~~~\Phi \to \phi\eta,~~~\Lambda \to \Lambda (e\eta)^2. 
\end{eqnarray}
As a result, the EYMH equations then depend only on the
dimensionless parameters $\alpha=M_{W}/(e M_{Pl})$,
$\beta=M_{H}/M_{W}$ and on $\Lambda$.

Within the above ansatz, the reduced EYMH action can be expressed as
\begin{eqnarray}
\label{action} S = \int  ~dr~dt \Big [  
\sigma m'-\alpha^2 \sigma (\omega'^2 N+\frac{(\omega^2-1)^2}{2r^2}
+\frac{1}{2}\phi'^2r^2N+\omega^2 \phi^2+V(\phi)r^2 )
\Big],
\end{eqnarray}
where the prime denotes the derivative with respect to the radial
coordinate $r$ and $V(\phi)=\beta^2(\phi^2-1)^2/4$, with $\beta^2=\lambda/e^2$.
 
The EYMH equations  reduce to the following system of four
non-linear differential equations
\begin{eqnarray}
\label{e1}
m'&=&\alpha^2(\omega'^2N+\frac{(\omega^2-1)^2}{2r^2}+\frac{1}{2}\phi'^2r^2N+\omega^2 \phi^2+V(\phi)r^2 ),
~~
\sigma'=-\frac{2\sigma}{r}\alpha^2(\omega'^2+\frac{1}{2}\phi'^2r^2),
\\
\label{e3}
(N\sigma \omega')'&=&\sigma \omega \big(\frac{(\omega^2-1)}{r^2}+ \phi^2 \big),
~~
(N\sigma r^2 \phi')'=\sigma(2  \omega^2 \phi+r^2\frac{\partial V}{\partial \phi}).
\end{eqnarray}

In the absence of the Higgs field, 
one  defines a set of  dimensionless  variables by performing 
the following rescalings 
$r \to (\sqrt{4\pi G}/e) r,~\Lambda \to (e^2/4 \pi G) \Lambda$
and $m \to (eG/\sqrt{4\pi G}) m$.
The equations of the EYM-$\Lambda$ system model then read
\begin{eqnarray}
\label{eym-eqs}
m'=\omega'^2N+\frac{(\omega^2-1)^2}{2r^2},
~~~
\sigma'=-\frac{2}{r}\sigma \omega'^2 ,
~~~
(N\sigma \omega')'=\sigma \frac{\omega(\omega^2-1)}{r^2}~.
 \end{eqnarray}
\subsection{Known solutions}
Several explicit  solutions  of the above equations are well known. 
The Schwarzschild-de Sitter solution corresponds to  
\begin{eqnarray}
\label{SdS}
\omega(r)=\pm 1,~~~~~\sigma(r)=1,~~~\phi(r)=0~~{\rm with} ~~v=0,~~
N(r)=1-\frac{2M}{r}-\frac{\Lambda r^2}{3},
\end{eqnarray}
and describes a black hole inside a cosmological
horizon as long as $N(r)$ has two positive zeros, $i.e.$ 
$M< 1/3 \sqrt{\Lambda}$.
The Nariai solution is found for $M= 1/3 \sqrt{\Lambda}$, 
in which case the function $N(r)$ has a double
zero at $r_c=1/\sqrt{\Lambda}$, corresponding to the position
of an extremal cosmological horizon.
 
The second solution  corresponds to an embedded U(1) configuration
with 
\begin{eqnarray}
\label{RNdS}
\omega(r)=0,~~\sigma(r)=1,~~\phi(r)=1,~~
N(r)=1-\frac{2M}{r}+\frac{\alpha^2}{r^2}-\frac{\Lambda r^2}{3}
\end{eqnarray}
and describes a particular parametrisation of the 
(magnetic-) Reissner-Nordstr\"om-dS (RNdS) black hole.
Since this solution proves to be important in understanding the
features of the  nonabelian configurations,
we present here a brief review of its properties
(see $e.g.$ \cite{Romans:1991nq}, \cite{Brill:1993tw}, \cite{Astefanesei:2003gw}
for more details).

For $\Lambda>0$, up to 
four zeros of $N(r)$ can exist. Three of the four zeros correspond to horizons
since the first zero has always negative value and thus has no physical meaning.
The two inner horizons $r_-$, $r_+$ with $r_- \le r_+$ correspond
to the well known Cauchy, respectively event horizon of the
Reissner-Nordstr\"om solution, while the third outer horizon
$r_c > r_+$ is the cosmological horizon.
Extremal black hole solutions - like in the AF case - are also
possible. Then, we have $r_-=r_+=r_h$ with $N(r_h)=N'|_{r=r_h}=0$.
This leads to the equation:  
\begin{equation}
\Lambda r_h^4 -r_h^2+\alpha^2=0 \ ,
\end{equation}
which is solved by 
\begin{equation}
\label{xhds}
r_{h/c}=\frac{1}{\sqrt{2\Lambda}}\sqrt{1\pm\sqrt{1-4\alpha^2\Lambda}} \ \
\ \ \ {\rm for} \ \ \  \frac{1}{4\alpha^2} \ge \Lambda > 0 \ .
\end{equation}
Obviously, the appearance of horizons
in dS space is restricted by $\alpha^2 \le \frac{1}{4\Lambda}$.
The corresponding mass of the extremal solution is given by:
\begin{equation}
M_{ext}=\frac{2}{3}\frac{\alpha^2}{r_h}+\frac{r_h}{3} \ .
\end{equation}
As the parameter $M$ increases (relative to $\alpha$, with $M>0$), the outer
black hole and cosmological horizons move closer together. The charged
Nariai solution is obtained when these horizons coincide at $%
r_{h}=r_{c}$; this is the largest charged asymptotically dS black
hole. For a given $\alpha$, the charged Nariai black hole has the maximal mass.
 
For completeness, we present also another solution which 
is less relevant to the situation discussed in this paper.
This solution is found for a particular value of the cosmological constant
$\Lambda=3/(16 \pi G)$ and a vanishing Higgs field $\phi=\beta=0$.
It reads 
\begin{eqnarray}
\label{einstein-univ}
ds^2=\frac{dr^2}{1-\frac{2}{3}\Lambda r^2}+r^2 d \Omega^2-dt^2,~~~~
\omega(r)=\sqrt{1-\frac{2}{3}\Lambda r^2},
\end{eqnarray}
corresponding to an $S^3\times R$ Einstein universe. 

\subsection{Boundary conditions for the EYMH model}
We want the generic line element (\ref{metric}) to describe a nonsingular,
asymptotically de Sitter spacetime outside a cosmological horizon located at $r=r_c>0$.
Here $N(r_c)=0$ is only a coordinate singularity where all curvature
invariants are finite. 
Outside the cosmological horizon $r$ and $t$ changes the character
($i.e.$  $r$ becomes a timelike  coordinate  for $r>r_c$).
A nonsingular extension across this null
surface can be found just as at the event horizon of a black hole.
The regularity assumption implies that all
curvature invariants at $r=r_c$ are finite.
Also, all matter functions and their first derivatives extend  smoothly
through the cosmological horizon, $e.g.$ in a similar way as the U(1) electric
potential $A_t=A_0+Q/r$ of a RNdS solution.

Similar to the case $\Lambda \leq 0$, it is natural to consider two types of configurations,
corresponding in the usual terminology to \emph{particle-like} and \emph{black hole}
solutions. 
  
For particle-like solutions, $r=0$ is a regular origin,  and we find
the following behaviour:
\begin{eqnarray}
\label{origin} 
\nonumber 
m(r)&=&\frac{1}{6}\alpha^2(12b^2+3a^2+2\beta^2)r^3+O(r^5),
~~
\sigma(r)=\sigma_0+\frac{1}{2}(a^2+8b^2)\sigma_0\alpha^2 r^2+O(r^3),
\\
\omega(r)&=&1-br^2+O(r^4),
~~ 
\phi(r)=ar+O(r^3),
\end{eqnarray}
where $a$ and $b$ are free parameters which are found by solving the
field equations.

\subsubsection{Expansion at the horizon}
The cosmological horizon is located at some finite value of $r$, where $N(r_c)=0$,
$\sigma(r_c)\neq 0$, the gauge potential and the scalar field taking some 
constant values.
The corresponding expansion as $r \to r_c$ is
\begin{eqnarray}
\label{cosm} 
\nonumber 
m(r)&=&\frac{r_c}{2}(1-\frac{\Lambda r_c^2}{3})+m_1(r-r_c)+O(r-r_c)^2,
~~~
\sigma(r)=\sigma_c+\sigma_1(r-r_c)+O(r-r_c)^2,
\\
\omega(r)&=&\omega_c+\omega_1(r-r_c)+O(r-r_c)^2,
~~~
\phi(r)=\phi_c+\phi_1(r-r_c)+O(r-r_c)^2,
\end{eqnarray}
where  
\begin{eqnarray}
\nonumber
m_1=\alpha^2(\frac{1}{2r_c^2}(\omega_c^2-1)^2+\omega_c^2\phi_c^2+V(\phi_c)r_c^2),
~~
\omega_1=\frac{1}{N_1}(\frac{\omega_c(\omega_c^2-1)}{r_c^2}+\omega_c\phi_c^2),
\\
\phi_1=\frac{1}{N_1r_c^2}(2\omega_h^2 \phi_h^2+r^2 \frac{\partial V}{\partial \phi}\Big|_{r_c}),
~~
\sigma_1=-\frac{2\sigma_c}{r_c}\alpha^2 (\omega_1^2+\frac{1}{2}\phi_1^2 r_h^2),
\end{eqnarray}
 $\sigma_c,~\omega_c,~\phi_c$ being arbitrary parameters and 
 $N_1=N'(r_c)=(1-2m_1-\Lambda r_c^2)/r_c$.
Note that  the condition $N'(r_c)<0$ should be satisfied, which imposes 
the following constraint
\begin{eqnarray}
\frac{1}{2r_c^2}(\omega_c^2-1)^2+\omega_c^2\phi_c^2+V(\phi_c)r_c^2>\frac{1-\Lambda r_c^2}{\alpha^2}.
\end{eqnarray}

We will consider also cosmological black hole solutions. 
These configurations possess an event horizon located at 
some intermediate value of the radial coordinate
$0<r_h<r_c$, all curvature
invariants being finite as $r \to r_h$.
As $r \to r_h$, the functions $m,\sigma,\omega$ and $\phi$ present an expansion
very similar to (\ref{cosm}) 
\begin{eqnarray}
\label{eh} 
\nonumber 
m(r)&=&\frac{r_h}{2}(1-\frac{\Lambda r_h^2}{3})+m'(r_h)+O(r-r_h)^2,
~~~
\sigma(r)=\sigma_h+\sigma'(r_h)(r-r_h)+O(r-r_h)^2,
\\
\omega(r)&=&\omega_h+\omega'(r_h)(r-r_h)+O(r-r_h)^2,
~~~
\phi(r)=\phi_h+\phi'(r_h)(r-r_h)+O(r-r_h)^2,
\end{eqnarray}
where  
\begin{eqnarray}
\nonumber
m'(r_h)=\alpha^2(\frac{1}{2r_h^2}(\omega_h^2-1)^2+\omega_h^2\phi_h^2+V(\phi_h)r_h^2),
~~
\omega'(r_h)=\frac{1}{N'(r_h)}(\frac{\omega_h(\omega_h^2-1)}{r_h^2}+\omega_h\phi_h^2),
\\
\phi'(r_h)=\frac{1}{N'(r_h)r_h^2}(2\omega_h^2 \phi_h^2+r^2 \frac{\partial V}{\partial \phi}\Big|_{r_h}),
~~
\sigma'(r_h)=-\frac{2\sigma_h}{r_h}\alpha^2 (\omega'(r_h)^2+\frac{1}{2}\phi'(r_h)^2 r_h^2),
\end{eqnarray}
 $\sigma_h,~\omega_h,~\phi_h$ being arbitrary parameters and 
 $N'(r_h)=(1-2m'(r_h)-\Lambda r_h^2)/r_h$.
The obvious condition $N'(r_h)>0$  imposes 
the  constraint
\begin{eqnarray}
\frac{1}{2r_h^2}(\omega_h^2-1)^2+\omega_h^2\phi_h^2+V(\phi_h)r_h^2<\frac{1-\Lambda r_h^2}{\alpha^2}.
\end{eqnarray}
Both the event  and the cosmological
horizon have their own  surface gravity $\kappa$ 
given by
\begin{eqnarray}
\nonumber
\kappa^2_{h,c}=- \frac{1}{4}g^{tt}g^{rr}(\partial_r g_{tt})^2\Big|_{r=r_h,r_c}~~,
\end{eqnarray} 
the associated Hawking temperature being $T_H=|\kappa|/(2\pi)$.

\subsubsection{Expansion at infinity}
The analysis of the field equations as $r\to\infty$ 
implies a more complicated picture
as compared to the AF or asymptotically AdS case, since the 
cosmological constant enters in a nontrivial way the solutions' expression
at infinity.

We suppose that the Higgs scalar approaches asymptotically its vev, while the 
magnetic gauge potential $w(r)$ vanishes.
This assures the absence of supplementary 
contributions  to the cosmological vacuum energy (apart from $\Lambda$) as $r \to \infty$,
while the gauge field approaches asymptotically the U(1)-Dirac monopole field.
The asymptotic expression of the metric functions $m(r),~\sigma(r)$ and the matter 
functions $w(r)$ and $\phi(r)$ is found by finding an approximate
solution of the field equations (\ref{e1})-(\ref{e3}) 
for these boundary conditions.

Here it is instructive to work with dimensionfull variables, the 
corresponding expression after considering the
rescaling (\ref{rescaling}) being straightforward.

A systematic analysis of the matter field equations
reveals that
a positive cosmological constant sets a mass bound for the
gauge sector,
 $M_{b}=\sqrt{\Lambda/12}$.
 The  expression of the gauge field as $r \to \infty$ for $M_W<M_b$, which leads to 
 finite mass solutions is
\begin{eqnarray}
\label{as3} w(r) \sim c_1 r^{k_1},~~~ 
{\rm with}~~~
k_1=-\frac{1}{2}\left(1+\sqrt{1-M_W^2/M_b^2}\right)~.
\end{eqnarray}
This contrasts with the exponential decay found in an asymptotically flat spacetime.

For $M_W> M_b$, the large $r$ behaviour of the gauge field is
\begin{eqnarray}
\label{as4}
 w(r) \sim  c_3 r^{-1/2} \sin \Psi_1(r)~,
~~~{\rm with}~~~\Psi_1(r)=(P_1\log r+c_4),~~P_1=\frac{1}{2}\sqrt{ M_W^2/M_b^2-1},
\end{eqnarray}
which we will find leads to a logarithmic divergence in the asymptotic expression
of the mass function $m(r)$.

The analysis of the scalar field asymptotics
is standard. Strominger's mass bound 
\cite{Strominger:2001pn}
($i.e.$ the dS version
of the Breitenlohner-Freedman
bound \cite{Breitenlohner:1982bm})
 is $M_{S}^2= 3\Lambda/4$  
and separates the infinite energy solutions from solutions which may
present a finite mass (this would depend also on the gauge field
behaviour). For small enough values of the Higgs field mass,
$M_H<M_{S}$ the scalar field decay leading to a finite asymptotic
value of $m(r)$ is
\begin{eqnarray}
\label{as1} 
\phi(r) \sim
\eta+c_2r^{k_2}~~,
{\rm with}~~~
k_2=-\frac{3}{2}
\left( 1+
\sqrt{1- M_H^2/M_{S}^2}
~\right)~.
\end{eqnarray}
We found numerical evidence for the
existence of a secondary branch of solutions decaying as 
\begin{eqnarray}
\label{as11} 
\phi(r) \sim
\eta+c_2r^{\tilde k_2}~,
{\rm with}~~~
\tilde k_2=-\frac{3}{2}
\left( 1-
\sqrt{1- M_H^2/M_{S}^2}
~\right).
\end{eqnarray}
However, this decay leads to a divergent value of the function $m(r)$
as $r \to \infty$.

For a Higgs mass  exceeding Strominger's bound,
the scalar field behaves asymptotically as
\begin{eqnarray}
\label{as2} 
\phi(r) \sim
\eta+c_5 r^{-3/2}\sin \Psi_2(r)~,
~~~{\rm with}~~~\Psi_2(r)=(P_2\log r+c_6),~~P_2=\frac{3}{2}\sqrt{ M_H^2/M_S^2-1},
\end{eqnarray} 
 (the constants $c_i$ which enter the above relations are free
parameters).

One may think that this bound may be circumvented 
by solutions with a vanishing Higgs potential.
However, 
by rewriting the Higgs field equation in the form
\begin{eqnarray}
\nonumber
\frac{1}{2}(N\sigma r^2 (\phi^2)')'&=&\sigma(N  r^2 \phi'^2+2w^2 \phi^2  
+r^2 \phi \frac{d V}{d \phi}),
\end{eqnarray}
and integrating it between the origin  and the 
cosmological horizon,
it can easily be proven that
no nontrivial solutions exist for $V(\phi)=0$ or for a convex potential.

Once we know the asymptotics of the matter fields,
the corresponding expression for the metric functions
 results straightforwardly from the equations (\ref{e1}).

 For $M_H<M_{S}$ and $M_W<M_b$ the mass of solutions tends to a constant 
 value as $r \to \infty$ and we find the asymptotic expressions
\begin{eqnarray}
\label{mass-reg} 
m(r) &\sim& 4\pi G \left(M
+(\eta^2-\frac{\Lambda}{3g^2})\frac{c_1^2 r^{2k_1+1}}{2k_1+1}
+(\frac{\lambda\eta^2}{2}-\frac{\Lambda k_2^2}{6})\frac{c_2^2 r^{2k_2+3}}{2k_2+1}
-\frac{1}{2g^2 r} \right)+\dots,\\
\label{s-reg} 
\log \sigma(r) &\sim& -8 \pi G(\frac{k_1^2c_1^2r^{2k_1-2}}
{2g^2(k_1-1)}+\frac{k_2c_2^2}{4}r^{2k_2})+\dots~.
\end{eqnarray}
In the case $M_W>M_b$ the metric function $m(r)$ 
gets a divergent contribution from the nonabelian field, whose
  leading order expansion is
\begin{eqnarray}
\label{mass-div-w} 
m^{YM}(r) \sim 4\pi G M_1-\frac{\pi G c_3^2}{P_1} \bigg(
(-2\Psi_1+\sin 2 \Psi_1)\eta^2
+\frac{\Lambda}{12 g^2}(8P_1\cos^2\Psi_1
\\
\nonumber
+2(1+4P_1^2)\Psi_1+(4P_1^2-1)\sin 2\Psi_1)\bigg) +\dots,
\end{eqnarray}
(note the presence in this expression of a divergent term originating in the kinetic Higgs term).

For a Higgs field mass exceeding the Strominger bound $M_H>M_{S}$,
we find a very similar  asymptotic form of the mass function $m(r)$,
presenting the same type of divergencies
\begin{eqnarray}
\label{mass-div-fi} 
m^{H}(r) \sim 4\pi G M_2 
-\frac{c_5^2\pi G}{4P_2}
\bigg(
\lambda \eta^2 (-2\Psi_2+\sin 2\Psi_2)
+\frac{\Lambda}{6}(24 P_2\cos^2\Psi_2
\\
\nonumber
+2(9+4P_2^2)\Psi_2
+(-9+4P_2^2)\sin2 \Psi_2)
\bigg) +\dots~.
\end{eqnarray}
Since the order of the cosmological constant is believed to be
much smaller than both the gauge and Higgs boson
mass, our results indicate that monopoles  in the universe
will always have a divergent mass function $m(r)$  as evaluated at timelike infinity.
Moreover, as discussed in Section 4,  
we could not find numerical solutions with  $M_H<M_{S}$ and $M_W<M_b$.

One should also study the asymptotic expression of the 
$\sigma$ function. Supposing $M_H>M_{S}$ and $M_W>M_b$ one finds
\begin{eqnarray}
\label{s-div-fi} 
\log \sigma (r) \sim 
\frac{\pi G}{3r^3}
\bigg[\frac{c_3^3}{9+4P_1^2} \left((1+4P_1^2)(9+4P_1^2)
+3(-3+4P_1^2)\cos2\Psi_1
-6P_1(5+4P_1^2)\sin 2\Psi_1\right)
\\
\nonumber
+c_5^2(9+4P_2^2-9\cos2\Psi_2-6P_2\sin2\Psi_2)\bigg].
\end{eqnarray}
The corresponding expressions for $M_H>M_{S}$ or 
 $M_W>M_b$ can easily be read from (\ref{s-div-fi}), (\ref{s-reg}).
 
The solutions with $M_H=M_{S}$,
 $M_W=M_b$ saturate these bounds and lead also to infinite mass configurations.
The EYMH equations lead to a matter fields expression
 \begin{eqnarray}
\label{sup1}
 w(r) \sim  \frac{c_7}{\sqrt{r}}+\frac{c_8\log r}{\sqrt{r}},~~~~
\phi(r) \sim
\eta +\frac{c_9}{\sqrt{r^3}}+\frac{c_{10}\log r}{\sqrt{r^3}},
\end{eqnarray}
(with $c_7,..,c_{10}$ real constants),
while for metric functions one can write $m(r)=m^{YM}(r)+m^{H}(r)$,
where
 \begin{eqnarray}
\label{sup2}
m^{YM}(r)&=&4 \pi G
\left(
-\frac{\Lambda}{36c_8g^2}(c_7-2c_8+c_8\log r)^3
+\eta^2(c_7^2\log r+c_7c_8\log r+\frac{c_8^3}{3}\log^3 r)
\right)+\dots
\\
\nonumber
m^{H}(r)&=&4 \pi G\left(
-\frac{\Lambda}{216c_{10}}(-2c_{10}+3c_9+3c_{10}\log r)^3
+\frac{\lambda}{6}\eta^2\log r
(3c_9^2+3c_9c_{10}\log r+c_{10}\log^2 r)\right)+\dots,
\end{eqnarray} 
 and
 \begin{eqnarray}
\label{sup3} 
\log \sigma(r)=
-\frac{2\pi G}{27 g^2r^3}
\left(9c_7-30c_7c_8 +26c_8^2+
3c_8 \log r(6c_7-10c_8+3c_8\log r)\right)
\\
\nonumber
-\frac{\pi G}{3r^3}\left(2c_{10}-6c_9c_{10}+9c_9^2
+3c_{10}\log r(-2c_{10}+6c_9
+3c_{10}\log r)\right)+\dots~.
\end{eqnarray} 
One can see that even for a logarithmic diverging
mass function,
the spacetime is still asymptotically dS, presenting 
the same conformal structure at infinity
as the vacuum dS solution, since
$g_{tt} \sim
-1+\Lambda r^2/3+O(\log r/r)$.
Also, the Ricci scalar stays finite as $r \to \infty$, $R \to 4\Lambda$.

Solutions with $\Lambda<0$ approaching at infinity the AdS background
despite the presence of a divergent  ADM mass have been considered 
by various authors in the last years, mostly for a scalar field matter content.
Restricting to the case of models with nonabelian matter fields in the bulk,
we mention the SU(2) hairy black holes in \cite{Radu:2004xp},
the family of globally regular solutions in 
D=4, ${\cal N}=4$ gauged supergravity in \cite{Chamseddine:2004xu},
and the self-gravitating Yang-monopoles in \cite{Gibbons:2006wd}.
The situation
for $\Lambda>0$ is less studied. However, 
it is natural to expect the existence of solutions with similar features for a 
dS background too,
the EYMH monopoles being a rather complicate case.

\subsection{Boundary conditions for the EYM model}
The field equations imply
the following behaviour for $r \to 0$ in terms of two real parameters $b,~\sigma_0$
\begin{eqnarray}
\label{origin2}
w(r) = 1-br^2 +O(r^4),~~m(r)=2b^2r^3+O(r^4),
~~\sigma(r)=\sigma_0+4b^2\sigma_0 r^2+O(r^4).
\end{eqnarray}
The corresponding expansion near the comological horizon is
\begin{eqnarray}
\label{expansion}
&m(r)=\frac{r_c}{2}\left(1-\frac{\Lambda r_c^2}{3}\right)+\frac{(\omega_c^2-1)^2}{2 r_c^2}(r-r_c),~~
\omega(r)=\omega_c+\frac{r_c\omega_c(\omega_c^2-1)}
{(1-\Lambda r_c^2)r_h^2-(\omega_h^2-1)^2}(r-r_c),
\\
\nonumber
&\sigma(r)=\sigma_c-\frac{2\sigma_h}{r_c} \omega'(r_h)^2 (r-r_c)+O(r-r_h)^2,
\end{eqnarray}
where $w_c,~\sigma_c$ are real parameters.

When discussing the pure EYM system with $\Lambda>0$, there
are no restrictions on the asymptotic value of the gauge potential
\cite{Volkov:1996qj}.
The field equations imply
the following expansion at large $r$
\begin{eqnarray}
\nonumber
&&m(r)= 
M+\left(\frac {\Lambda C_1^{2}}{3}-\frac {1}{2} (
\omega _{0 }^{2}-1) ^{2}\right)\frac {1}{r} +O\left( \frac {1}{r^{2}} \right),
~~
\omega (r) = \omega _{0 } +\frac {C_1}{r}
+O\left( \frac {1}{r^{2}} \right),
\\
\label{asympt}
&&\sigma(r)=1- \frac{C_1^2}{r^4}+O\left( \frac {1}{r^5}\right),
\end{eqnarray}
where $\omega_0$, $M$ and $C_1$   are constants
 determined by numerical calculations. 
 
 For cosmological black hole solutions having a regular event horizon at $r=r_h>0$,
we find the following expansion near the event horizon
\begin{eqnarray}
\label{expansion2}
&m(r)=\frac{r_h}{2}\left(1-\frac{\Lambda r_h^2}{3}\right)+\frac{(\omega_h^2-k)^2}{2 r_h^2}(r-r_h),~~
\omega(r)=\omega_h+\frac{r_h\omega_h(\omega_h^2-1)}
{(1-\Lambda r_h^2)r_h^2-(\omega_h^2-1)^2}(r-r_h),
\\
\nonumber
&\sigma(r)=\sigma_h-\frac{2\sigma_h}{r_h} \omega'(r_h)^2 (r-r_h)+O(r-r_h)^2,
\end{eqnarray}
 with $w_h, ~\sigma_h$ real parameters.

\section{Counterterm method and conserved charges}
\subsection{General formalism}

The computation of the conserved charges including mass 
in an asymptotically dS 
spacetime is
 a  difficult task.  This is due to the absence of the spatial 
 infinity and  of a globally timelike
Killing vector in  such a spacetime. 
One prescription to compute conserved charges in asymptotically dS was developed
by Abbott and Deser \cite{Abbott:1981ff}. 
In this perturbative approach one  considers the 
deviation of metric from pure dS space which is the vacuum of the theory
and measure the energy of fluctuations to find the mass. 

In \cite{Balasubramanian:2001nb}, a novel prescription
was proposed, the  obstacles mentioned above being
avoided by computing the quasilocal
tensor of Brown and York (augmented by the AdS/CFT  inspired  counterterms), 
on the Euclidean surfaces at future/past timelike infinity $\mathcal{I}^{\pm}$. 
The conserved charge associated with the Killing vector $\partial/\partial t$
- now spacelike outside the cosmological horizon-
was interpreted as the conserved mass. 
This allows also a discussion of the thermodynamics
of the asymptotically dS solutions outside the event horizon, 
the boundary counterterms regularising the (tree-level)  gravitation action.
The efficiency of this approach has been demonstrated in a broad range of examples.

A thorough discussion of the general formalism has been given $e.g.$
in \cite{Clarkson:2004yp}, and so we only recapitulate it here.
In this approach one starts by considering the general path integral 
\begin{equation}
\left\langle g_{2},\Psi _{2},S_{2}|g_{1},\Psi _{1},S_{1}\right\rangle =\int
D \left[ g,\Psi \right] \exp \left( iI\left[ g,\Psi \right] \right),
\label{PI1}
\end{equation}
which represents the amplitude to go from a state with metric and matter
fields $\left[ g_{1},\Psi _{1}\right] $ on a surface $S_{1}$ to a state with
metric and matter fields $\left[ g_{2},\Psi _{2}\right] $ on a surface $%
S_{2} $. \ The quantity $D\left[ g,\Psi \right] $ is a measure on the space
of all field configurations and $I\left[ g,\Psi \right] $ is the action
taken over all fields having the given values on the surfaces $S_{1}$ and $%
S_{2}$.
For asymptotically dS spacetimes we replace the surfaces $S_{1},S_{2}$ with
histories $H_{1},H_{2}$\ that have spacelike unit normals and are surfaces
that form the timelike boundaries of a given spatial region. The amplitude (%
\ref{PI1})  describes quantum correlations between differing histories 
$\left[g_{1},\Psi _{1}\right] $ and $\left[ g_{2},\Psi _{2}\right] $\ of metrics
and matter fields, with the modulus squared of the amplitude yielding the
correlation between two histories.  

In this approach, the initial action (\ref{lag0}) is supplemented by the  
 boundary counterterm action $I_{ct}$
depending only on geometric invariants of these spacelike surfaces. 
$I_{ct}$ regularizes the gravitational
action and the boundary stress tensor.  
In four dimensions, the counterterm expression is
(in this section we do not consider rescaled quantities; also 
to conform with standard conventions 
in the literature on this subject, we note $\Lambda=3/\ell^2$)
\begin{eqnarray}
\label{actionct}
I_{ct} =-\frac{1}{8\pi G}\int_{\mathcal{\partial M}^{\pm }} d^{3}x\sqrt{h}
( -\frac{2}{\ell }%
+\frac{\ell   }{2 }\mathsf{R}
)
\end{eqnarray}%
with $\mathsf{R}$ the curvature of the induced metric $h_{ij}$
and $\int_{\mathcal{\partial M}^{\pm }}$ indicates the sum of the
integral over the early and late time boundaries. 
In what follows, to simplify the picture,
 we will consider the  $\mathcal{I}^{+}$ boundary only,
dropping the $^{\pm }$ indices (similar results hold for $\mathcal{I}^{-}$).

The boundary metric can be written, at least locally, in a ADM-like general form 
\begin{equation}
ds^{  2}=h_{ij}^{  }d {x}^{  i}d {x}^{  j}=N_{t}^{ 2}dt ^{2}+\sigma _{ab}^{ }\left( d\psi ^{  a}+N^{  a}dt
\right) \left( d\psi ^{ b}+N^{  b}dt \right) ,  \label{hmetric}
\end{equation}%
where $N_{t }$ and $N^{a}$ are the lapse function and the shift vector
respectively and the $\psi ^{a}$ are the intrinsic coordinates on the closed
surfaces $\Sigma $.  
Varying the action with respect to the boundary metric $h_{ij}$, we find the
boundary stress-energy tensor for gravity 
\begin{equation}
T_{ij}=\frac{2}{\sqrt{h^{  }}} \frac{\delta I}{\delta h_{ij} },
\end{equation}%
the corresponding expression  for four dimensions being~\cite{Ghezelbash:2002ab}
\begin{eqnarray}
\label{bst}
T_{ ij}=
 -\frac 1{8\pi G}\left( K_{ij}-K h_{ij}-\frac
2l\left[ h_{ij}+\frac{\ell^2}2\left( \mathsf{R}_{ij}-\frac 12\mathsf{R}h_{ij}\right)
\right] \right),
\end{eqnarray}
where  $K_{ij}$ and $\mathsf{R}_{ij}$ are the extrinsic curvature  and the 
Ricci tensor of the boundary metric, respectively. 

In this
approach, the conserved  quantity associated with a Killing vector $\xi
^{ i }$ on the $\mathcal{I}^{+}$ boundary  is  given by 
\begin{equation}
\label{Qcons}
\mathfrak{Q_{\xi}}{}^{  }=\oint_{\Sigma ^{ }}d^{n}\phi ^{  }\sqrt{%
\sigma ^{  }}n^{  i}T_{ij}^{  }\xi ^{  j} ,  
\end{equation}%
where $n^{ i}$ is an outward-pointing unit vector, normal to surfaces of
constant $\tau$.  Physically, this means that a collection of
observers, on the hypersurface with the induced metric $h_{ij}$, would all
measure the same value of $\mathfrak{Q}_{\xi }$ provided this surface has an
isometry generated by $\xi ^{i}$. 

If $\partial /\partial t$ is a Killing vector on $\Sigma$, then 
the conserved mass is defined to be the conserved quantity 
$\mathfrak{M}$ associated with it.

A tree-level evaluation of the path integral (\ref{PI1}) for the 
EYM(-H) system may be carried out along the lines
described in ref. \cite{Ghezelbash:2002ab} for the vacuum case 
(see also the discussion in \cite{Astefanesei:2003gw} for the 
case of a gravitating U(1) field). 
Since the action is in general
negative definite near past and future infinity (outside of a cosmological
horizon), we analytically continue the coordinate orthogonal to the
histories to complex values by an anticlockwise $\pi/2$-rotation of
the axis normal to them.   This generally imposes, from the regularity
conditions, a periodicity $\tilde{\beta}$ of this coordinate, which is the analogue
of the Hawking temperature outside the cosmological horizon.

This renders the action pure imaginary, yielding a convergent path integral 
\begin{equation}
Z^{\prime }=\int e^{+\hat{I}}  
\label{Partitionaction}
\end{equation}%
since $\hat{I}<0$. 
In the semi-classical approximation this will lead to $%
\ln Z^{\prime }=+I_{cl}$.

 For a canonical ensemble with fixed
temperature  we can write 
\begin{equation}  \label{W}
F= \mathfrak{M}-TS,
\end{equation}%
where $F$ is the  canonical  potential and $T=1/\tilde{\beta}$. For a
converging partition function, we have $F=I_{cl}/\tilde{\beta}$ and thus we find for
the entropy of the system 
\begin{equation}  
\label{S}
S =\tilde{\beta} \mathfrak{M}-I_{cl}.
\end{equation}

\subsection{Spherically symmetric EYM(-H) solutions}
\subsubsection{The boundary stress tensor}
The application of the general formalism to
spherically symmetric EYM(-H) solutions discussed in the previous sections 
is straightforward.
Working outside the
 cosmological  horizon, 
we set following \cite{Ghezelbash:2002ab} $r=\tau$  
and rewrite the metric ansatz (\ref{metric})  as
\begin{eqnarray}
\label{m1}
ds^2=-f(\tau) d\tau^2+\frac{\sigma^2(\tau)}{f(\tau)}dt^2+\tau^2 d\Omega^2~,
\end{eqnarray}
where 
\begin{eqnarray}
f(\tau)=\left(\frac{\tau^2}{\ell^2}+\frac{2m(\tau)}{\tau}-1\right)^{-1}~.
\end{eqnarray}
We choose $\partial {\mathcal M}$ to be a three surface of fixed $\tau>r_c$,
which gives $n_{i}=1/\sqrt{f(\tau)}\delta_{\tau i}$).
The extrinsic curvature $K_{ij}=h_i^k\nabla _{k}n_j$ has the nonvanishing components 
\begin{eqnarray}
\label{Kij}
K_{\theta \theta}=-\frac{\tau}{\sqrt{f(\tau)}},~~
K_{\varphi \varphi}=-\frac{\tau \sin^2 \theta}{\sqrt{f(\tau)}},~~
K_{tt}=\frac{\sigma (\sigma f'-2f \sigma')}{2f^{5/2}},
\end{eqnarray}
the corresponding expression for the trace of the extrinsic curvature being
(in this section we do not consider rescaled quantities;
also the prime here denotes the derivative with respect to $\tau$)  
\begin{eqnarray}
\label{trace}
K=-\frac{2}{\tau \sigma}+\frac{\sigma f'-2f \sigma'}{2f^{3/2}\sigma}.
\end{eqnarray}
For EYM solutions or EYMH configurations with  $M_H<M_{S}$ and $M_W<M_b$
we find
from (\ref{bst}) that the nonvanishing components of 
the boundary stress-tensor are 
\begin{eqnarray}
\label{BD4}
\nonumber
T_{\theta}^{\theta}=T_{\varphi}^{\varphi} =\frac{1}{8 \pi G}
\frac{\ell M}{\tau^3}+O\left(\frac{1}{\tau^4}\right),~~~
T_{t}^t=-\frac{1}{4 \pi G}
\frac{M \ell}{ \tau^3}+O\left(\frac{1}{\tau^4}\right).
\end{eqnarray} 
The mass of these solutions measured at the far future boundary
of dS space, as computed from (\ref{Qcons}), is
\begin{eqnarray}
\label{Mct}
\mathfrak{M}=-M,
\end{eqnarray}
where $M$ is the parameter entering the asymptotic expansions (\ref{mass-reg}), (\ref{asympt}).

As discussed in the next section, all EYM solutions we found 
have $M>0$ (both black holes and particle like solutions).  (Note that
we didn't find any EYMH solutions with finite mass.) 
Thus  $\mathfrak{M}$ is negative, consistent with the
expectation \cite{Balasubramanian:2001nb} that pure dS spacetime has the
largest mass for a singularity-free spacetime $\mathfrak{M}=0$. 
If there is
a CFT dual to a magnetic monopole, this mass translates into the energy of the dual CFT
living on a Euclidean cylinder $R\times S^{2}$.

The boundary stress-tensor of the EYMH solutions with $M_H>M_{S}$ or $M_W>M_b$
(the only configurations we could find numerically is)
\begin{eqnarray}
\label{BD4_2}
\nonumber
T_{\theta}^{\theta}=T_{\varphi}^{\varphi} =\frac{1}{8 \pi G}
\frac{g_1(\tau)}{\tau^3}+O\left(\frac{\log \tau}{\tau^4}\right),~~~
T_{t}^t=-\frac{1}{4 \pi G}
\frac{g_1(\tau)}{ \tau^3}+O\left(\frac{\log \tau}{\tau^4}\right),
\end{eqnarray} 
where $g_1(\tau)$ can be read from (\ref{mass-div-fi}), (\ref{mass-div-w})
presenting a  logarithmic  divergence in $\tau$. 
This implies a divergent mass as computed from (\ref{Qcons}).

\subsubsection{The magnetic charge}
The computation of the magnetic charge in the absence of a spacelike infinity is 
also difficult.
However,  as  similar problems  appear  already in the abelian case, we may use 
the solution proposed in \cite{Astefanesei:2003gw},
which generalises the methods of ref.~%
\cite{Jolien} to dS case.
Working again outside the cosmological horizon, 
one defines the usual 't Hooft field strength tensor
\begin{eqnarray}
\label{hooft}
{\mathcal F}_{\mu \nu}=Tr\{ \hat{\Phi}F_{\mu \nu}-\frac{i}{2}D_\mu \hat{\Phi}D_\nu \hat{\Phi}  \},
\end{eqnarray}
where   $\hat{\Phi}=\Phi/|\Phi|$.
The induced metric $h_{\mu \nu
}=g_{\mu \nu }+u_{\mu }u_{\nu }$ projects the electromagnetic field on a
specific slice of the foliation,
where the vector 
$u^{\nu }=  \frac{1}{\sqrt{\left| f\left( \tau\right)\right| }} \delta^\nu_t$ 
is normal to the induced metric $h_{\mu \nu }$
and $n^{i}= \sqrt{\left| f\left( \tau\right) \right| }\delta^i_\tau $
is the (timelike) unit vector normal to hypersurface $\Sigma $.
The magnetic field with respect to a slice
$ \tau=const.$ is $B_{i}=\frac{1}{2}\epsilon_{ijk}{\mathcal F}_{jk } $ and the  charge
density at future infinity is 
\[
\rho _{\mathbf{Q}}=\sqrt{\sigma ^{  }}n^{  i}B_{i}^{  }, 
\]%
Integrating the charge density over the hypersurface $\Sigma$, we
obtain the total magnetic charge at future/past infinity  
\begin{eqnarray}
\label{m-charge}
\mathbf{Q}_m&=&\frac{1}{4\pi}\oint_{\infty}dS_{\mu}\frac{1}{2}\epsilon_{\mu \nu \alpha}
Tr\{\hat{\Phi}F_{\nu \alpha}\}.
\end{eqnarray}
For the particular case of spherically symmetric solutions one finds $\mathbf{Q}_m=1$
\footnote{However, we expect all AF configurations to present dS counterparts, in 
particular the axially symmetric monopole-antimonopole solutions, 
with a vanishing net magnetic charge
$\mathbf{Q}_m=0$ \cite{Kleihaus:2000hx}. 
Such solutions exist also for $\Lambda<0$ \cite{Radu:2004ys}.}.
The interpretation of the magnetic charge is analogous to that for $\mathfrak{M}$  noted above.
 $\mathbf{Q_m}$ is the charge measured
by a collection of observers  following  a given history; its conservation  
implies that observers following a different history would measure
the same value of the magnetic charge 
\footnote{This definition has nothing to do with 
the finiteness of the mass-energy of monopole solutions.
}.

The magnetic charge of the EYM solutions is evaluated at future/past infinity according to
(see e.g.\cite{Bjoraker:2000qd})
\begin{eqnarray} 
\label{SU2}
\mathbf{Q_m}=\frac{1}{4 \pi}\int dS_k \sqrt{-g} \tilde{F}^{kt}=
Q_m  \frac{\tau_3}{2},
\end{eqnarray}
where $Q_m=(1-\omega_{0 }^2)$. 

\subsubsection{The entropy and action}
It can also be proven that the entropy of the monopole solutions
associated with the cosmological event horizon is one quarter of
the cosmological event horizon area, as expected. 

Here we remark that, similar to the abelian theory \cite{Hawking:1995ap}, 
the partition function 
is evaluated taking the magnetic charges as a boundary condition.
Fixing the gauge potential fixes the magnetic charge directly.

By integrating the Killing identity
$\nabla^a\nabla_b K_a=R_{bc}K^c,$ 
for the Killing field $K^a=\delta^a_t$ (which gives
$R_t^t=-(\sqrt{-g}g^{tt}g^{\tau \tau} g_{tt,\tau})'/\sqrt{-g})$), 
together with the Einstein equation
$R_t^t=(R-2\Lambda )/2-8\pi G T_t^t$,
it is possible to isolate the bulk action contribution at $r \to \infty$ and on the 
cosmological event horizon.
Here we use the observation that, 
for monopole solutions $(A_t=0)$, the term $T_t^t$  exactly cancels the matter field
lagrangean in the bulk action
 $\mathcal{L}_{M}$.
 The integration is from the cosmological horizon out to some fixed $\tau$ that will be 
 sent  to infinity. 
Therefore we shall work in the "upper patch" outside 
of the cosmological horizon.
The divergent contribution given by the surface integral term at infinity in 
$R_t^t$ is also cancelled by 
$I_{\rm{surface}}+I_{ct}$ and we arrive at the simple finite expression
\begin{eqnarray}
\label{itot}
I_{cl}=\frac{\tilde{\beta}}{2}(M+r_c^3/\ell^2).
\end{eqnarray}
Since $\partial/\partial t$ is a Killing vector, we may
periodically identify it and assign a value for $\tilde{\beta}$.
If we analytically continue $t \to it$, we obtain a metric
of signature $(-2,2)$. 
The submanifold of signature $\left( -,-\right) $ described by
the $(t,\tau )$ coordinates will have a conical singularity at $\tau =\tau
_{+}$ unless the $t$-coordinate is periodically identified with period 
\begin{equation}
\tilde{\beta}_{H}=\left| \frac{4\pi }{(-\sigma^2(r)N^{\prime }(r))}\right| _{r=r_c}=
\left| \frac{-\sigma^2(r)f^{\prime }(\tau )}{4\pi  f^{2}(\tau )}\right| _{\tau
=r _{c}}^{-1}  \label{betaH}
\end{equation}%
This is the analogue of the Hawking temperature outside of the cosmological
horizon.
The relation $S=A_c/4$
results straightforwardly from (\ref{S}).

The overall picture for black hole solutions is, however, 
much more complicated than in the particle-like case.
This is because both black hole and dS space produce thermal radiation due to quantum effects. 
If these temperatures are different
(which is the generic case, see Figure 6), then the energy flows from the hotter horizon
to the cooler one and the black hole will gain or lose mass.

In deriving these relations we supposed that
the mass function $m(\tau)$ approaches a finite
value as $\tau \to \infty$.
This is the case for the monopole solutions in the EYM theory.
However, we could not find EYMH solutions with this 
feature and they are unlikely to exist. 
For EYMH configurations, the constant $M$ in (\ref{itot}) is formally infinite,
which leads to a divergent action, too.

For AdS solutions, 
 in some cases it is still  possible to obtain a finite mass by
allowing the regularising counterterms to depend not only on
the boundary metric  but
also on the matter fields on the boundary 
\cite{Taylor-Robinson:2000xw,Hollands:2005wt}.
We expect that this approach would apply also to dS EYMH monopoles,
yielding a finite mass/action to these solutions (see $e.g.$ the 
case of dS Goldstone solutions \cite{Brihaye:2005qr}).

Also, as remarked in \cite{Batrachenko:2004fd},
while the counterterms  are necessary to
render both the action and energy finite,
the thermodynamics is not affected by changing the counterterm
prescription (basically both $I_{cl}$ and $\tilde{\beta} \mathfrak{M}$
are shifted with the same amount).
Therefore we expect  the entropy of the EYMH monopoles,
evaluated outside the cosmological horizon to
satisfy also the generic relation $S=A_c/4$.

\subsection{Remarks on the boundary CFT}
As in the AdS/CFT correspondence, the metric of the manifold on which the 
\textit{putative} dual Euclidean CFT resides is defined by 
\[
\gamma _{ij}=\lim_{\tau\rightarrow \infty }\frac{\ell^2}{ \tau ^{2}} h_{ij}. 
\]%
The dual field theory's stress tensor $\tau _{k}^{i}$, is related to the
boundary stress tensor by the rescaling \cite{Myers:1999qn} 
\[
\sqrt{\gamma }\gamma ^{ij}\tau _{jk}=\lim_{\tau \rightarrow \infty }\sqrt{h}%
h^{ik}T_{jk}. 
\]%
For the bulk metric ansatz (\ref{metric}),
the geometry of the manifold on which the dual CFT resides is given by the
 cylinder metric 
\begin{eqnarray}
ds^{2}=\gamma _{ij}dx^{i}dx^{j}=dt^{2}+ {\ell^{2}}d\Omega^{2}.
\end{eqnarray}
The CFT's stress tensor is 
\begin{eqnarray}
\tau _{j}^{i}=\frac{1}{8\pi G} M (-3u^{i}u_{j}+\delta
_{j}^{i}),  \label{staticstress}
\end{eqnarray}
where $u^{i}=\delta _{t}^{i}$. This tensor is finite, covariantly conserved
and manifestly traceless, as expected from the  dS/CFT correspondence, since even dimensional
bulk theories are dual to odd dimensional CFTs which have a
vanishing trace anomaly. 

One can see that a divegent $\mathfrak{M}$ in the bulk would imply a divergent value for the
energy of the dual CFT.

\subsection{Quantities inside cosmological horizon}
Following \cite{Gibbons:1977mu}, one may also define
a total mass
$\mathbf{M}_c$
inside the cosmological
horizon.
This can be done by integrating the Killing identity
$\nabla^\mu\nabla_\nu K_\mu=R_{\nu\rho}K^\rho,$
 for the Killing field $K=\partial/\partial t$
on a spacelike hypersurface $\Sigma$ from the origin to
$r_c$ to get the Smarr-type formula
\begin{eqnarray}
\label{Smarr1}
\mathbf{M}_c \equiv \frac{1}{4 \pi G}\int \nabla_\mu K_\nu d\Sigma^{\mu \nu}=
\frac{1}{4 \pi G}\int _{V_c}\Lambda K_\mu d\Sigma^\mu+
\int_{V_c}(2T_{\mu \nu}-Tg_{\mu \nu})K^\mu d\Sigma ^\nu,
\end{eqnarray}
where one integrates between the origin and the cosmological horizon.
It is natural to identify the left-hand side as the total mass within the cosmological
horizon. 
The first term on the r.h.s. is the contribution of $\Lambda$ to the total mass within the
cosmological constant, while the last one is the 
contribution of the matter fields.
For the metric ansatz (\ref{metric}),
$\mathbf{M}_c$ can also be rewritten as $\mathbf{M}_c=- \kappa_cA_c/4\pi G=
-r_c^2\sigma(r_c) N'(r_c)/2G$,
where $\kappa_c,~A_c$ are the cosmological horizon surface gravity and
area, respectively.

A relation  similar to (\ref{Smarr1}) can also be written for black hole solutions
\begin{eqnarray}
\label{Smarr2}
\mathbf{M}_c \equiv \frac{1}{4 \pi G}\int \nabla_\mu K_\nu d\Sigma^{\mu \nu}=
\frac{\kappa_hA_h}{4\pi G}+
\frac{1}{4 \pi G}\int_{V_1} \Lambda K_\mu d\Sigma^\mu+
\int_{V_1}(2T_{\mu \nu}-Tg_{\mu \nu})K^\mu d\Sigma ^\nu.
\end{eqnarray}
One should notice that $\mathbf{M}_c$ in the above relations is
different from $\mathfrak{M}$, since the former keeps track only of the mass within 
the cosmological horizon, whereas the latter has contributions also from outside.

Also, the EYMH monopole solutions we found have  always  a finite 
$\mathbf{M}_c$, while  $\mathfrak{M}$ generically diverges.
\section{Numerical results} 

Although an analytic or approximate solution appears to be
intractable, we present in this section numerical arguments that the
known EYMH solutions can be extended to include a positive cosmological
constant. 
Apart from that, we discuss also monopole solutions of the EYM model,
emphasizing their properties at timelike infinity.

To integrate the field equations, we used in all cases the differential
equation solver COLSYS which involves a Newton-Raphson method
\cite{COLSYS}.
 The case $\Lambda > 0$ leads to the occurrence of a cosmological
 horizon at $r = r_c$ with $N(r_c) = 0$.
Similar to the U(1) case,
the cosmological horizon radius $r_c$ is a function of $\Lambda$.
Numerically, we have implemented a supplementary equation for the ``function''
$\Lambda$ with $\Lambda'=0$.
Solving the set of equations
fixing $r_c$ then gives us the solutions for a numerically determined
value $\Lambda=const.$. 
Using this trick, we can impose two further
boundary conditions for the ``function'' $\Lambda$, one being effectively a condition
on $m$ at $r_c$.
Our procedure then has been to first solve the
equations on the interval $[0:r_c]$ fixing $r_c$ and then solve
the equations on the interval $[r_c:\infty)$. In all our numerical
computations it then turned out that the solutions on $[0:r_c]$ and
$[r_c:\infty)$ can be combined to a continuous solution on $[0:\infty)$.

\subsection{Magnetic monopole solutions in EYMH theory}
The system of equations depends in this case on three parameters 
$(\Lambda$, $\alpha$, $\beta)$ in the case of 
gravitating monopoles and additionally on $r_h$ in the case of non-abelian
black holes.

Using the initial conditions at the origin/event horizon
and on the cosmological horizon,  
the equations of motion (\ref{e1})-(\ref{e3}) were solved
varying $\Lambda$ for a range of values of the coupling
parameter $\alpha$ and several values of $\beta$.

We here present our results
for $\beta = 0.1$ and remark that the results
are qualitatively similar for other values of $\beta$ we considered.


As a general feature of both particle-like and black hole
solutions, we note that
while a negative cosmological constant
exerts an additional pressure on configurations,
causing their
typical radius to become thinner \cite{Lugo:1999fm},
a positive $\Lambda$ has the opposite effect, causing the typical solution
radius to expand beyond the value it would have
in asymptotically flat space.

Hence, for small values of $\Lambda$, 
the solutions in the region $r<r_c$ resemble the AF solutions, being surrounded 
by a cosmological horizon and approach
dS geometry in the asymptotic region.

Also, as $\alpha$ increases, the cosmological horizon shrinks in size.
The dS configurations are generally not confined inside the cosmological horizon,
with all variables and their first derivatives extending smoothly
through the cosmological horizon.

  Note that the bounds for finite energy solutions 
$M_H < M_S$ and $M_W < M_b$ after the rescalings 
(\ref{rescaling}) read $2\beta < \sqrt{3 \Lambda}$ and $\sqrt{12} < \sqrt{\Lambda}$.
 
\subsubsection{Solutions with a regular origin}

\begin{figure}
\epsfysize=10cm
\epsffile{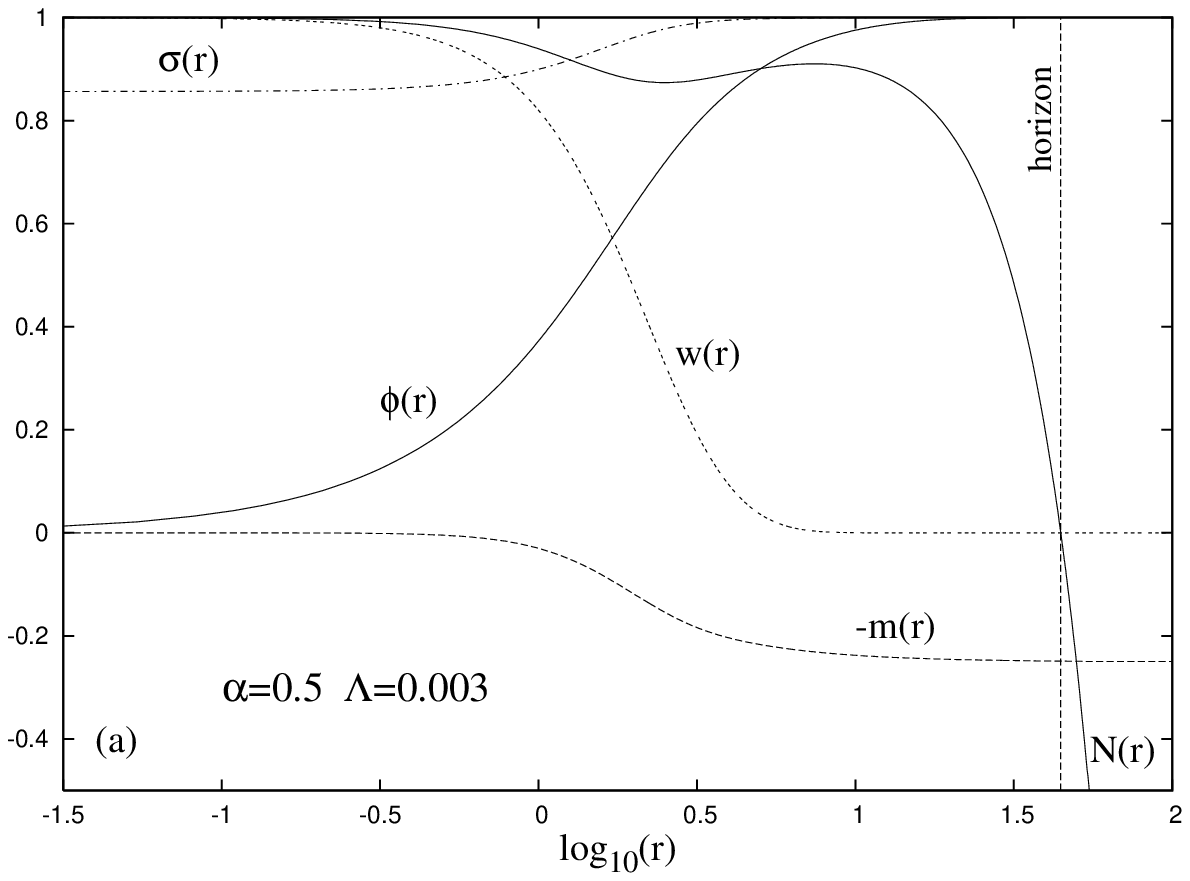}
\epsfysize=10cm
\epsffile{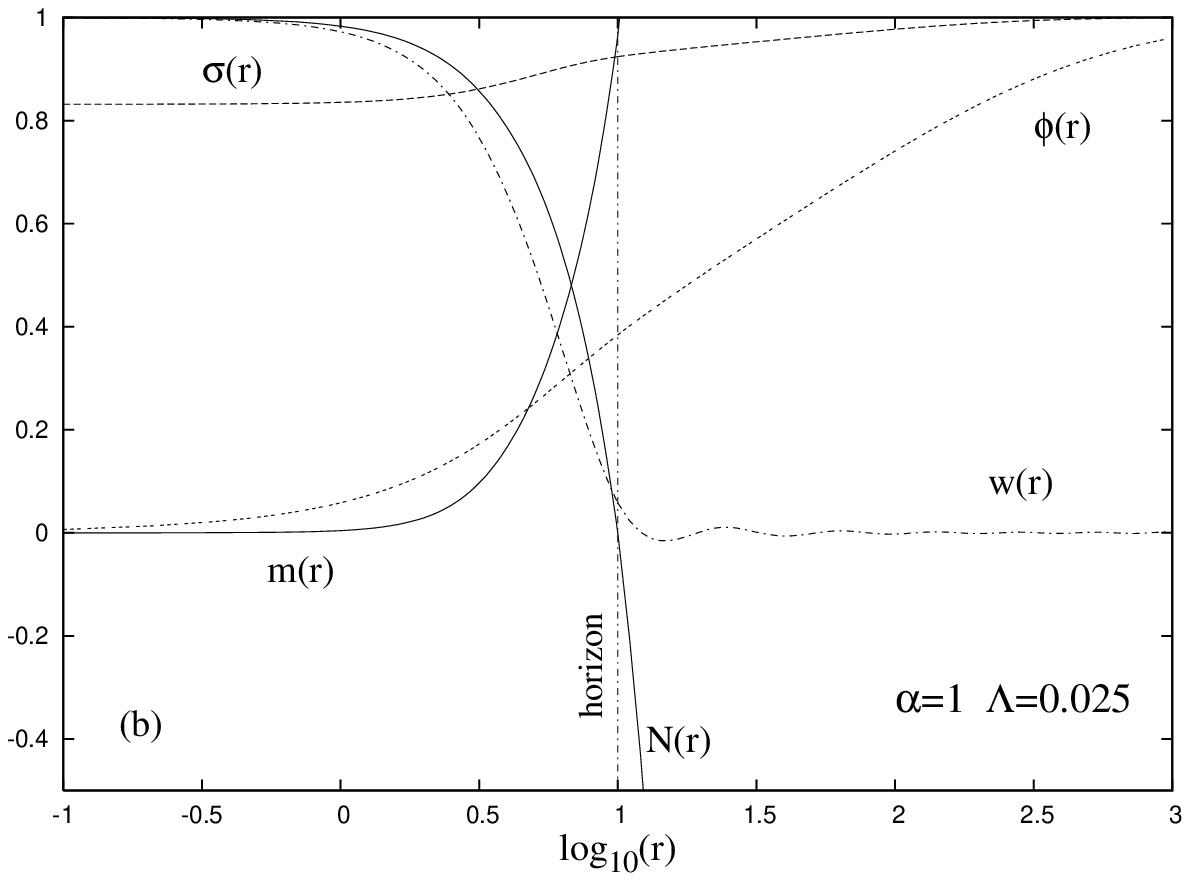}
\vskip 1cm
\caption{\label{eymh-reg} 
The profiles of two typical gravitating EYMH monopole solutions 
with a regular origin are plotted as a function of $r$ for $\alpha=1$. In (a), we show
a solution on the first branch for $\alpha=0.5$, in (b) a solution on the second branch for $\alpha=1.0$.
Here and in figures  \ref{eym8}, \ref{eym-2nodes} and \ref{eym-bh}
we have indicated also the position of the cosmological
event horizon.
}
\end{figure}

 When $\Lambda$ is increased from zero,
 while keeping $\alpha$, $\beta$ fixed,
a branch of dS solutions emerges from the corresponding asymptotically
flat
configurations. 
This branch ends at
a maximal value $\Lambda_{max}$.
The value of $\Lambda_{max}$ depends on the parameters $\alpha$,
$\beta$.
For example, for solutions with
 $\beta=0.1$ in a  fixed dS background ($\alpha=0$),
we find  $\Lambda_{max}\approx 0.035$.
The value of $\Lambda_{max}$ varies only little with
$\alpha$ and $\beta$,
e.g.  for $\alpha=1$, $\beta=0.1$
we find $\Lambda_{max}\approx 0.033$.
 The profile of a typical first-branch solution is presented in Figure \ref{eymh-reg}(a).

A second branch of  solutions  always 
appears at $\Lambda_{max}$
 (such that at $\Lambda_{max}$ the two branches merge at one solution),
 extending  backwards in $\Lambda$ to a zero
value  of the cosmological constant.
The trivial solution
 $\phi(r < \infty)\equiv 0$, $w(r < \infty)\equiv 1$
 with $\phi(r=\infty)=1$, $w(r=\infty)=0$  is approached
 (the solution thus
has an infinite slope at $r=\infty$) in this limit. 
This can be understood as follows:
since we have rescaled $\Lambda \to \Lambda/e^2 \eta^2$,
the limit $\Lambda \to 0$ on the first branch
corresponds to asymptotically flat space with $\eta \neq 0$.
On the second branch, however, the limit
$\Lambda \to 0$ corresponds to asymptotically de
Sitter space with $\eta\to 0$, which then implies
the trivial solution. We give the profile of a typical second-branch solution in Figure \ref{eymh-reg}(b).

The main difference between the solutions on the two
branches is that the matter field functions attain their asymptotic
values much quicker for the solutions on the first branch. 

\begin{figure}
\epsfysize=9cm
\epsffile{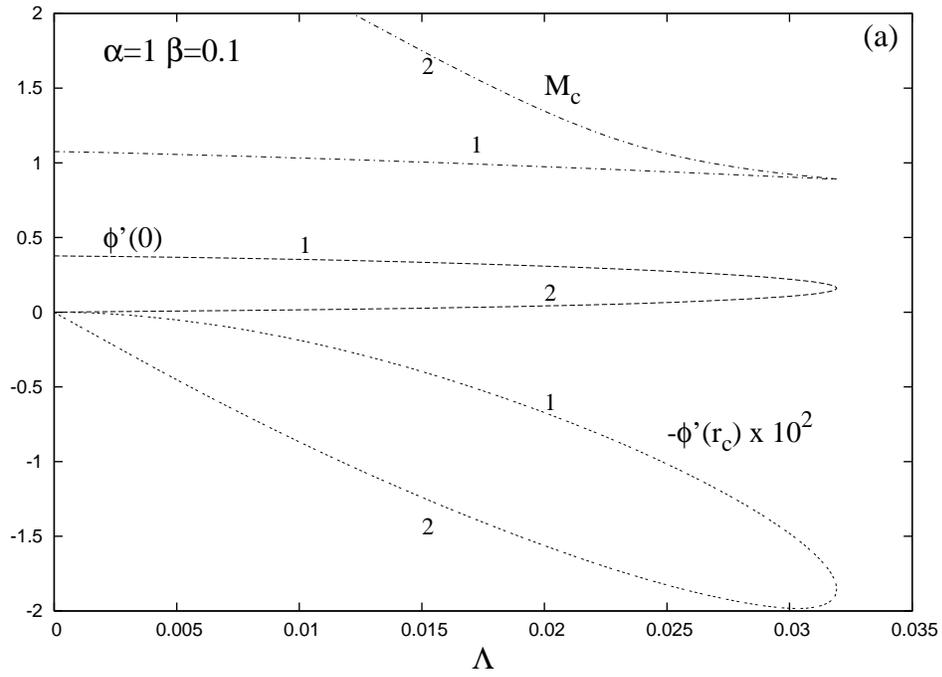}
\\
\\
\epsfysize=9cm
\epsffile{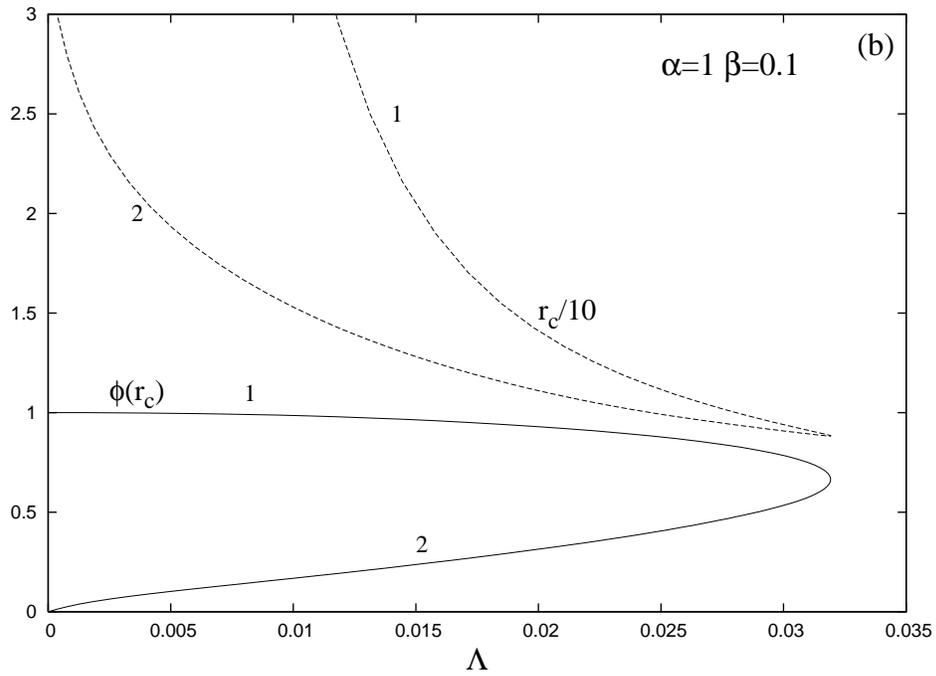}
\vskip 1cm
\caption{\label{eymh-spectrum}
  The dependence
of solutions' properties on the value of the cosmological constant
is plotted for particle-like monopole solutions.
The labels $1$ (resp. $2$)
refer to the first (resp. second) branch.
}
\end{figure}

This is shown in Figure \ref{eymh-spectrum}  where some numerical data is given as function
 of $\Lambda$ for the two branches.
The results in this figure are obtained for
$\alpha=1,~\beta=0.1$
but they remain qualitatively the same for all gravitating solutions
we have considered.
The maximal values of  $\Lambda$ are always below the critical value  
found for solutions
in fixed dS background, and
as a result the mass of our solutions measured at future/past infinity
always diverges. While we have
found solutions with $2\beta < \sqrt{3\Lambda}$ (finite contribution from
the Higgs field), all solutions have $\sqrt{\Lambda} < \sqrt{12}$. Thus
although the contribution from the Higgs field might be finite, the contribution
from the gauge field, however, leads always to an infinite mass. 
For small values of
$\Lambda$ (Figure 1a), 
the divergence in $m(r)$ will manifest itself only for very large values of $r$; 
however, this becomes obvious for large enough values of $\Lambda$ (see Figure 1b).
 
Note, though, that the mass $\mathbf{M}_c$
within the cosmological
horizon stays finite (see also Figure \ref{eymh-reg}) and that the mass  $\mathbf{M}_c$ doesn't tend
to zero in the limit of trivial solutions.
 
The existence of other disconnected branches of solutions
for $\Lambda>\Lambda_{max}$ appears unlikely.

\begin{figure}
\epsfysize=12cm
\epsffile{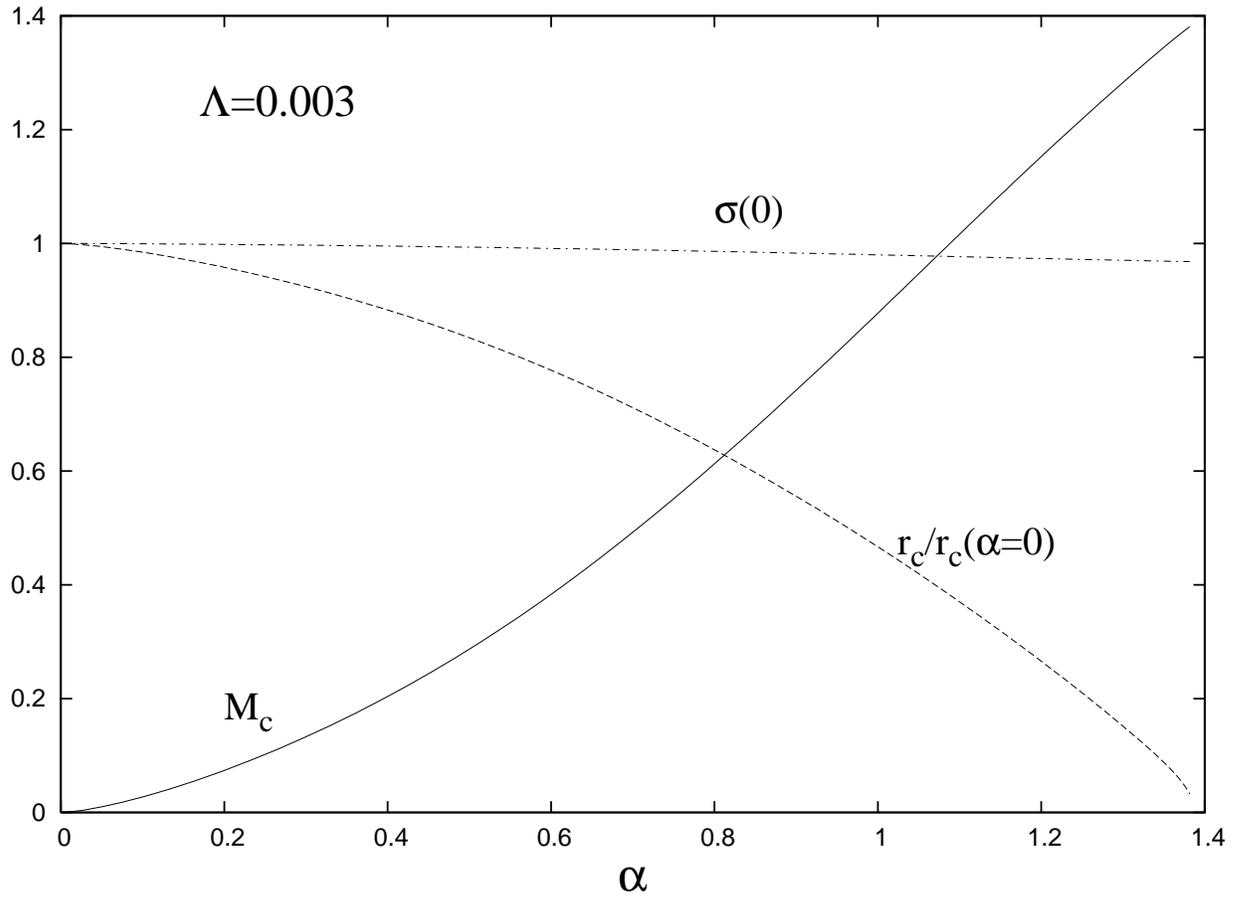}
\vskip 3cm
\caption{\label{fig1} 
The mass inside the cosmological horizon $\bf{M_c}$, the value of 
the metric function $\sigma$ at the origin, $\sigma(0)$,
and the value of the cosmological
horizon $r_c$ of the gravitating monopoles 
are given as functions of $\alpha$ for 
fundamental branch solutions with
$\Lambda = 0.003$.
}
\end{figure}

For a fixed $\Lambda<\Lambda_{max}$, our numerical analysis shows that regular
 solutions exist on a finite interval of the parameter $\alpha$ (which depends on $\beta$),
 $i.e.$ $\alpha \in [0, \alpha_{max}(\Lambda)]$. 
Correspondingly,
 the cosmological horizon depends slightly on $\alpha$~:
 $r_c(\alpha = \alpha_{max}) \leq r_c \leq r_{c}(\alpha = 0)$.
For $e.g.$  $\Lambda = 0.003$,
we find  $43.6 \leq  r_c \leq 44.7$, as  shown in  Figure~\ref{fig1}.
  

When the parameter $\alpha$ is increased, the local minimum
 of the metric function $N(r)$  becomes 
 deeper and deeper while at the same time the
 cosmological horizon decreases slightly. 
In the limit $\alpha \rightarrow \alpha_{max}$,
 the local minimum of $N$ approaches zero at
$r=r_{h}$ with $N(r=r_{h})=0$ and $N'(r)|_{r=r_{h}}=0$.
In fact this value is determined
 explicitely according to
 $$
      r_{h} = \frac{1}{\sqrt{2\Lambda}}
      \sqrt{1 - \sqrt{1 - 4 \alpha^2 \Lambda}}
  \ \ , \ \ {\rm for} \ \ \frac{1}{4 \alpha^2} > \Lambda \ \ >  0
  $$
.

  These results give strong evidence that 
  in the limit $\alpha \rightarrow \alpha_{max}(\Lambda)$
  the gravitating monopole
  solutions bifurcate with the branch of extremal
  RNdS
 solution on the interval $[r_{h}, \infty]$.
 The gravitating monopoles separate in this limit
into an interior region $r\in [0:r_{h}]$ with a smooth origin and a nontrivial YM field,
and an exterior region $r\in [r_h:\infty]$ where the solution
corresponds to an extremal RNdS solution with $w=0$, $\phi=1$ and thus
unit charge.
 As a consequence for $r$ close to and larger than
 the cosmological horizon $r_c$ 
  the metric functions $\sigma(r)$, $N(r)$
  are identical to ones of the RNdS solution.
 The mass value $m_{\infty}$
can then be determined explicitely by imposing 
the condition $N(r_c)=0$.
We also observe
that  in the limit $\alpha \rightarrow 0$ 
the value $\sigma(0)$ decreases considerably, while
the value $r_c$ tends to a finite (but $\Lambda-depending$)
  value. The solution corresponds to an hyperbolic monopole.

As stressed above, the value of the cosmological 
 horizon depends only slightly on $\alpha$.
 This is shown in Figure~\ref{eymh-spectrum}. 
We found also that
 $M_c$ and $\sigma(0)$ depend rather weakly on $\Lambda$.

\subsubsection{Black hole solutions}
\begin{figure}
\epsfysize=12cm
\epsffile{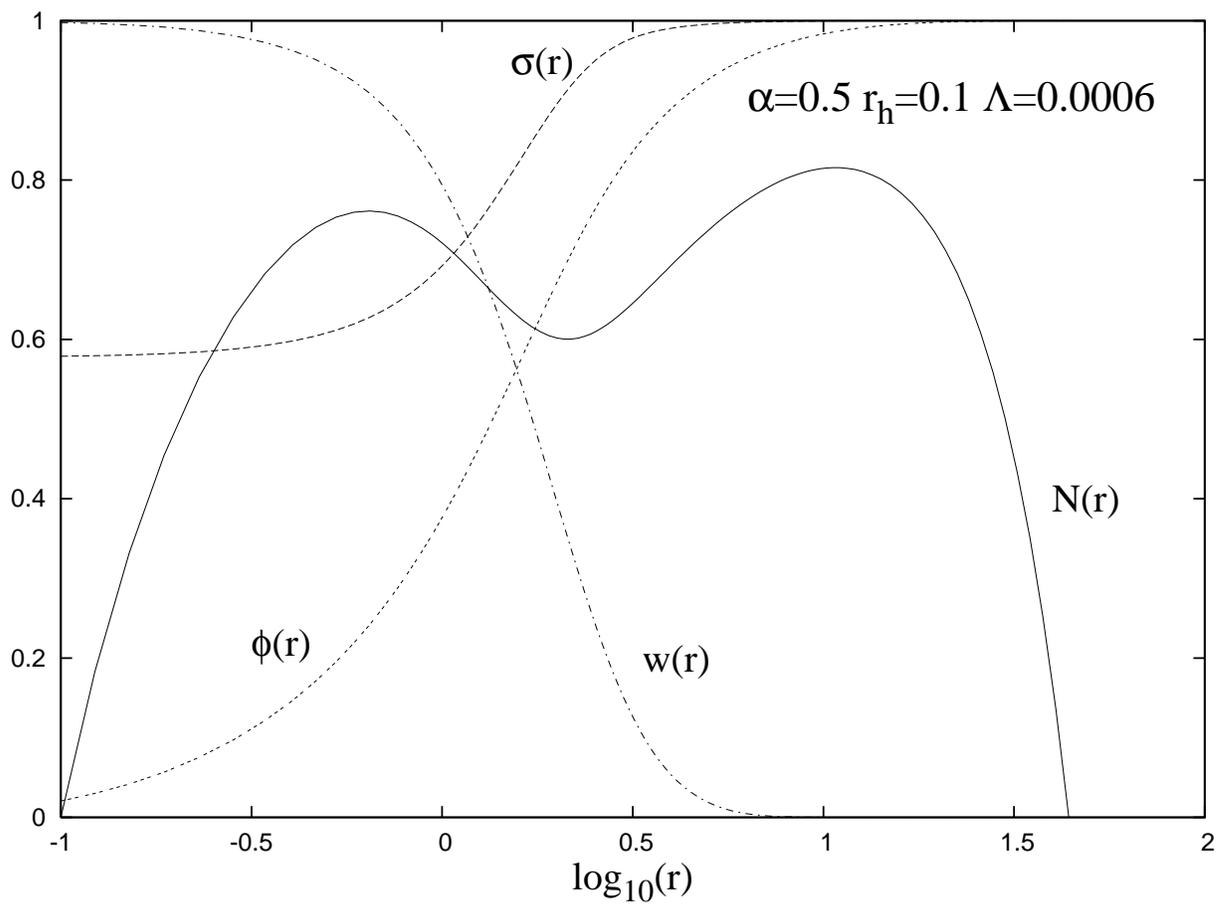}
\vskip 3cm
\caption{\label{fig5} 
The profiles of the functions of the non-abelian black holes are given
for $\alpha=0.5$, $r_h=0.1$ $\Lambda=0.0006$.
}
\end{figure}

Similarly to the globally regular solutions, 
non-abelian black holes solutions can be constructed
with $\Lambda > 0$. These solutions correspond to non-abelian
black holes sitting inside the core of de Sitter monopoles.
They have a double horizon: an event
horizon at $r = r_h$ and a cosmological horizon at  $r = r_c$.
The function $N(r)$ is zero at these two points. A generic  de Sitter
black hole solution is presented in Figure~\ref{fig5} for $\alpha=0.5$,
$\Lambda=0.0006$ and $r_h=0.1$.
The event horizon is fixed by hand while the cosmological horizon
depends on the choice of $\alpha$, $\Lambda$.
$r_c$ is then determined numerically.

\begin{figure}
\epsfysize=12cm
\epsffile{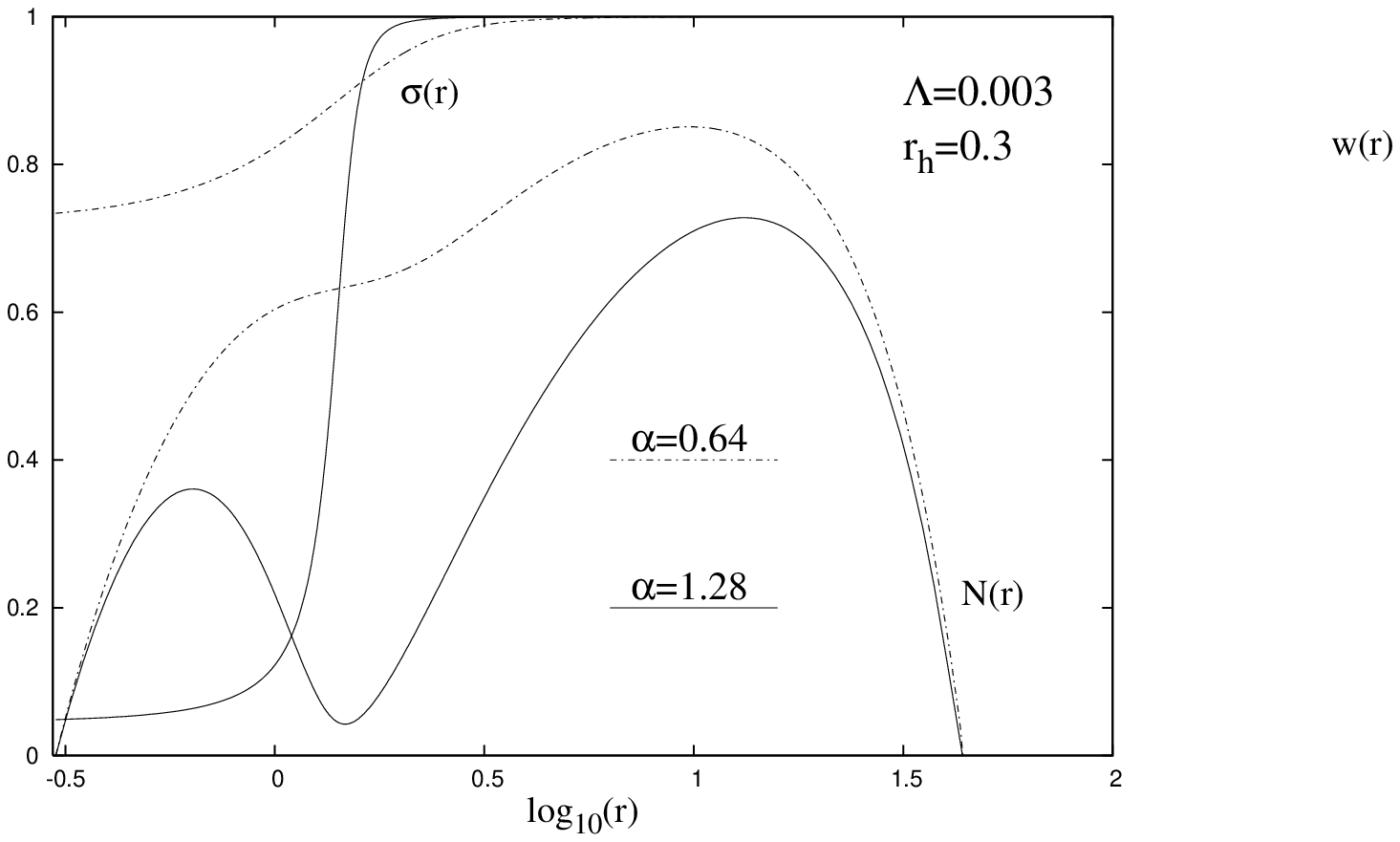}
\vskip 3cm
\caption{\label{fig6} 
The metric functions $N$ and $\sigma$ of two EYMH
black hole solutions with $r_h = 0.3$, $\Lambda = 0.003$ and
two different values of $\alpha$ are shown as a function of $r$.
}
\end{figure}

For fixed $r_h$, and sufficiently large $\alpha$
the  metric function $N(r)$ develops
a local minimum at $r = r_{min}$, $r_h < r_{min} < r_c$ 
and the value $N(r_{min})$
decreases and tends to zero for a maximal value of $\alpha$.
This is seen in Figure \ref{fig6} for $\Lambda = 0.003$
and $r_h = 0.3$.  Very similarly to the regular case,
in this critical limit, we end up with a solution presenting three
horizons. The intermediate one corresponds a double zero of the 
metric function $N(r)$ with $N(r=r_{min})=0$, $N'(r)|_{r=r_{min}}=0$.
On the interval $[r_{min}, \infty)$ the solution approaches
an extremal RNdS solution, while on the interval
$[r_h, r_{min}]$ the non-abelian black hole solution remains non trivial.

This phenomenon is illustrated in Figure~\ref{fig6} for $\Lambda = 0.003$
and $r_h = 0.3$.

\begin{figure}
\epsfysize=12cm
\epsffile{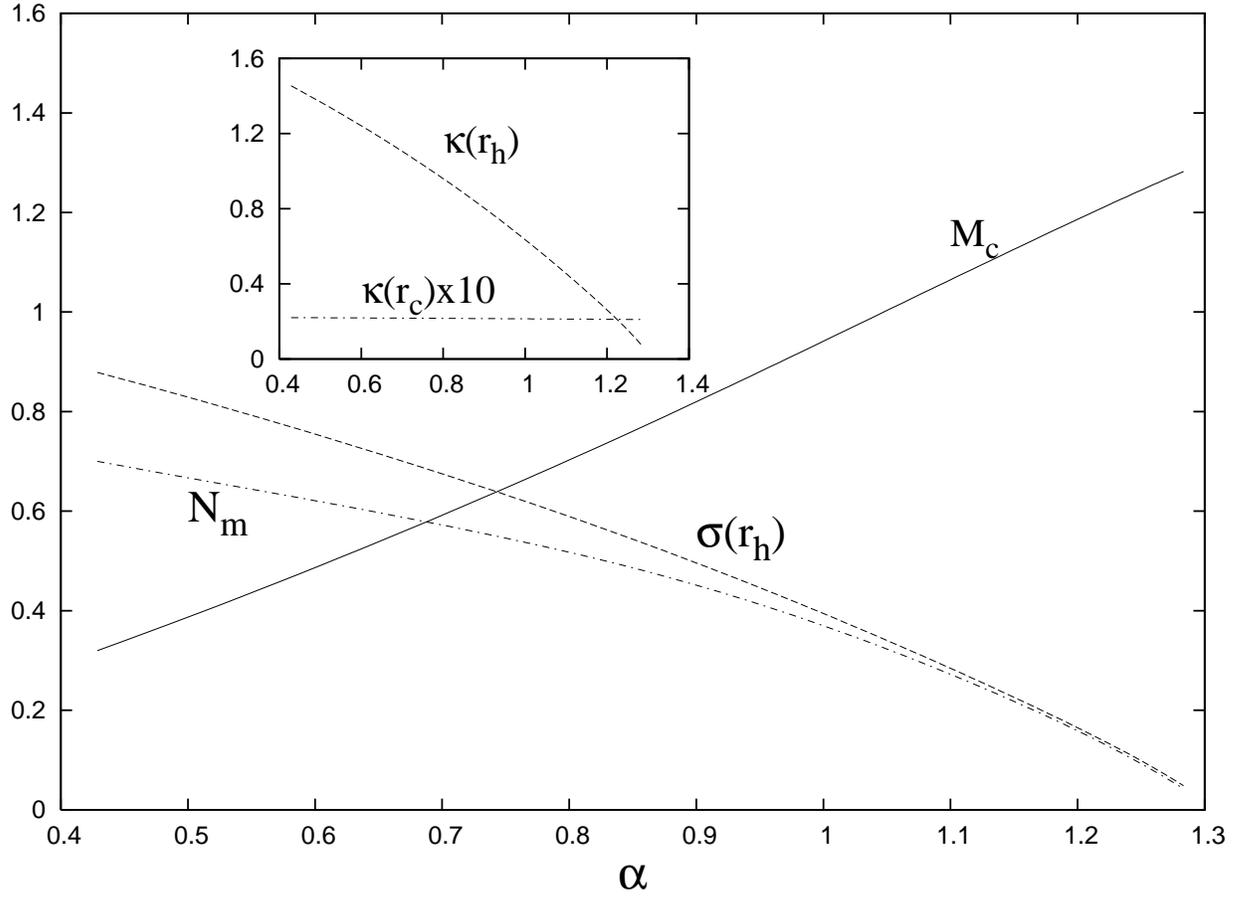}
\vskip 3cm
\caption{\label{fig7} 
 The surface gravity at the two horizons,
the  mass inside the cosmological horizon $\bf{M_c}$ and the minimal value $N_m$ of the metric function
$N$, are shown as functions of $\alpha$
for non-abelian EYMH black hole solutions with $r_h = 0.3$, $\Lambda = 0.003$.
}
\end{figure}

At each horizon of these black holes solutions, the surface gravity
(determining the entropy) can be computed. In terms of the
spherically symmetric ansatz, it is given by
$$
    \kappa^2 = \frac{1}{2} N'(r)  \sigma(r)\Big|_{r=r_c,r_h}~. 
$$
The various quantities characterising the black hole are
shown in Figure~\ref{fig7}. In particular, we see that the 
surface gravity at the inner event horizon decreases considerably
when increasing $\alpha$ and becomes very small (along
with $\sigma(r_h)$) in the limit $\alpha \rightarrow \alpha_{max}$.
In contrast,  the surface gravity at the cosmic horizon
varies only little with $\alpha$ (along with $\sigma(r_c) = 1$).

Similar to the case of solutions with regular origin,
the mass function $m(r)$ of the black hole solutions
also diverges at infinity, while the mass inside the cosmological horizon stays finite.
In order to demonstrate the peculiar behaviour of the Higgs
function $\phi(r)$ for $r \to \infty$, we have to 
construct solutions for large values of $\Lambda$. In Figure 7, we show our results
for $\Lambda = 0.033$. 
We especially show the value of
$r \phi'/(1-\phi)$. The asymptotic value of this particular combination
(which tends to some constant $\sim 0.75$ for the parameters chosen here) 
confirms clearly the behaviour (\ref{as11}).
Of course, this phenomenon is present 
for generic values of the parameters 
although not always easy to set up numerically. We were able to exhibit
this property for regular solutions as well.

\subsection{Magnetic monopoles in EYM theory} 
We have integrated the system of eqs. (\ref{eym-eqs})
for a range of $\Lambda$ by using the same techniques.
Both solutions with regular origin and cosmological
black holes have been considered.

All numerical solutions we found have $w^2\leq 1$,
which implies a nonzero node number of the gauge potential $w$.
This can be proven 
by using the sum rule \cite{Bjoraker:2000qd}
\begin{eqnarray}
-(\frac{N\sigma \omega'}{\omega})\Big|_{r_0}^{r_c}
=\int_{r_0}^{r_c}dr~\sigma \big(\frac{1-\omega^2}{r^2}+
N\frac{\omega'^2}{\omega^2}\big),
\end{eqnarray}
which follows directly from the YM equations 
($r_0$ here is $r_0=0$ for solutions with regular origin or $r_0=r_h$,
for black holes).
Suppose that $w(r)$ never vanishes and $w^2\leq 1$ for $r_0<r<r_c$.
Then l.h.s. of the above relation vanishes, while the
integrand of the  r.h.s.  is positive definite.
Therefore the gauge potential
of the nontrivial YM configurations with $w^2\leq 1$ must vanish at least once
in the region inside the cosmological horizon.

Thus for a given value of $\Lambda$, 
the solutions are indexed by the node number $k$.
Different from the AF case, however, the asymptotic value of the
gauge potential is not fixed
by finite energy considerations. Similar to $M$, $w_{0}$ appears
as a result of the numerical integration.

For small values of the cosmological constant $\lambda \ll 1/R_c^2$
($R_c$ corresponding to the typical solutions' radius),
the contribution $\Lambda r^2$ to the energy density is negligible.
For $r \ll R_c$, the solutions do not considerably deviate 
from the corresponding flat space configurations.
In the region $r>R_c$, the effects of $\Lambda$ become significant.

Several characteristic features of the two node particle-like solutions
of the EYM-$\Lambda$ system are plotted as a function of the cosmological
constant in Figure 8.
One can see that both the mass inside the cosmological horizon
and the mass evaluted at timelike infinity are finite. 
The next two plots show the profiles of typical EYM-$\Lambda$ solutions with regular origin 
(Figure 9) and cosmological black holes  (Figure 10).
In both cases, $w^2(\infty)\neq 1$, which implies the existence of a 
nonvanishing nonabelian magnetic charge defined according to
(\ref{SU2}).

One should remarks that the EYM theory presents solutions with dS asymptotics only for
values of the cosmological constant up to some $\Lambda_{max}<3/4$ \cite{Volkov:1996qj}.

As discussed in \cite{Lu:2003dm},
the dS EYM-SU(2) theory is a consistent truncation of $D=11$ 
supergravity via Kaluza-Klein dimensional reduction on a non-compact space,
the positive cosmological constant being  fixed by the SU(2) gauge coupling constant
$\Lambda=4e^2$. 
The ``internal'' space is a 
smooth hyperbolic seven-space  
written as a foliation of two three-spheres, 
on which the SU(2) Yang-Mills fields reside.
However, no solutions with dS asymptotics exist for this value of the cosmological
constant \cite{Breitenlohner:2004fp}.

\section{The nonabelian monopole in cosmological coordinates}
Although the line element (\ref{metric}) takes a simple form in a static
coordinate system, the expansion of the universe is not manifest.
Furthermore, the static coordinates $(t,r)$ break down at the Killing
horizons.
However, 
one can prove that the results in the previous sections
 remain qualitatively
unchanged when using a different parameterisation of dS spacetime.
A particularly interesting case is provided by the planar coordinates
(for which the spatial slices are flat) --- 
or cosmological or inflationary coordinates --- with a dS line-element
\[
ds^{2}=e^{2HT}(dR^{2}+R^{2}d\Omega^{2})-dT^{2},  
\]
(to conform with the standard notation in literature, we note here 
$\Lambda=3H^2$, \textit{i.e.} $H=1/\ell$).
The relation between the physics in a static coordinate system 
and in a planar coordinate system in the case of Einstein-U(1) theory
is discussed in \cite{Astefanesei:2003gw}.

Considering the general metric ansatz (\ref{metric}) 
and the matter fields form (\ref{A}), (\ref{higgs-form}), 
the solution in the cosmological form can be written as:
\beqs
ds^2&=&-V^{-2}dT^2+U^2e^{2H T}(dR^2+R^2d\Omega^2),
\nonumber
\\
A&=&\frac{1}{2e}\bigg[\omega(Re^{H T})\tau_1d\theta+(\cos\theta\tau_3+\omega
(Re^{Ht})\tau_2\sin\theta)d\varphi\bigg],~~~~
\Phi=\phi(Re^{Ht})\tau_3
\eeqs 
where $U$ and $V$ are functions of $\rho=Re^{H T}$. 
The coordinate transformation that relates (\ref{metric}) to the above metric form is:
\beqs
 U(\rho)&=&\frac{r(\rho)}{\rho},~~~~~~t=T-\int\frac{H r dr}{\sigma(r)N(r)\sqrt{\sigma(r)^2 N(r)+H^2r^2}},\\
V^{-2}(\rho)&=&\sigma^2(r(\rho))\left(\frac{U+\rho U'}{U}\right)^2,
\eeqs
while $r(\rho)$ is determined implicitly by:
\beqs
\int\frac{\sigma(r)dr}{r\sqrt{\sigma^2(r)N(r)+r^2H^2}}=\ln\rho
\label{Rho}.
\eeqs
As an example, let us consider the $U(1)$ solution with:
\beqs
\omega(r)=0,~~~~~\sigma(r)=1,~~~~~~\phi(r)=1,~~~~~
N(r)=1-\frac{2M}{r}+\frac{\alpha^2}{r^2}-Hr^2
\eeqs
Then we can easily integrate (\ref{Rho}) and we obtain:
\beqs
U(\rho)=1+\frac{M}{\rho}+\frac{M^2-\alpha^2}{4\rho^2},~~~~
V(\rho)=\frac{1+\frac{M}{\rho}+\frac{M^2-\alpha^2}{4\rho^2}}{1-\frac{M^2-\alpha^2}{4\rho^2}}.
\eeqs
which is readily seen to be the RNdS solution in cosmological coordinates. 
The extremality condition can be written in this case as $|\alpha|=M$.

One advantage of using the planar coordinates is that the expansion of 
the universe is now manifest: the 
cosmological expansion chart corresponds to $H>0$, 
while the contracting chart is given by $H<0$. However, 
this comes with the price that the manifest time-translation 
symmetry is broken while the charts that cover the horizons are 
highly distorted. The boundary geometry ${\cal I}^{\pm}$ is now 
approached for large $T$ and the boundary topology is $R^3$ 
(written above in spherical coordinates).
 
Let us notice that $\rho=Re^{H T}$ is unchanged under the transformations:
\beqs
T\ra& T+a, ~~~~~~ R\ra e^{-Ha}R,
\eeqs
which means that the geometry in the cosmological ansatz is preserved. 
The first term in the above transformation generates the time 
translations in the bulk, while the second term corresponds to 
a scale transformation of the radial coordinate on the boundary. 
There exists a Killing vector associated with this symmetry and it can be written as:
\beqs
\xi&=&-H R \frac{\partial}{\partial R}+\frac{\partial}{\partial T}~.
\eeqs
Similar to the (electro-)vacuum case, 
the norm of this Killing vector will vanish precisely where 
$N(r)=0$, that is, at the horizons $r=r_h,~r_c$ as determined 
in static coordinates.

It is straightforward to compute the boundary stress-tensor in 
cosmological coordinates. Choosing the boundary $\partial{\cal M}$ 
to be the three-surface given by a fixed value of $T$, the normal 
to this surface is $n_{\mu}=1/V\delta_{T\mu}$ we find the following 
components of the extrinsic curvature:
\beqs
K_{RR}&=&-\frac{HU^2a^2}{\sigma},~~~~~K_{\theta\theta}=-
\frac{HU^2a^2r^2}{\sigma},~~~~~K_{\varphi\varphi}=-\frac{HU^2a^2r^2\sin^2\theta}{\sigma}
\eeqs
that is, $K_{ij}=-\frac{H}{\sigma}h_{ij}$, where $h_{ij}$ 
is the induced metric on $\partial{\cal M}$ and we denoted $a(T)=e^{HT}$. 

The components of the boundary stress-tensor in cosmological 
coordinates are then given by:
\beqs
8\pi GT_{RR}&=&\frac{2U^2a^2}{\sigma^2}\left(\frac{H}{\sigma}-
\frac{1}{\ell}\right)-\frac{2\ell U_{,R}}{rU}\left(1+\frac{RU_{,R}}
{2U}\right)
=\frac{2M\ell}{R^3a}+O(a^{-2})
\nonumber
\\
8\pi GT_{\theta\theta}&=&\frac{2U^2a^2R^2}{\sigma^2}
\left(\frac{H}{\sigma}-\frac{1}{\ell}\right)-\ell r^2
\bigg[-\left(\frac{U_{,R}}{U}\right)^2+
\frac{U_{,R}}{RU}+\frac{U_{,RR}}{U}\bigg]
=-\frac{2M\ell}{Ra}+O(a^{-2})
\nonumber
\\
8\pi G T_{\varphi\varphi}&=&\frac{2U^2 a^2 R^2
\sin^2\theta}{\sigma^2}\left(\frac{H}{\sigma}
-\frac{1}{\ell}\right)-\ell R^2\sin^2\theta\bigg[
-\left(\frac{U_{,R}}{U}\right)^2+\frac{U_{,R}}{rU}+\frac{U_{,RR}}{U}\bigg]
=-\frac{2M \ell \sin^2\theta}{R a}+O(a^{-2})\nonumber
\eeqs
where we assumed an expansion of the form $U(\rho)=1+\frac{M}{\rho}+O(\frac{1}{\rho^2})$.
Notice that in the limit $T\ra\infty$ 
the first terms on the right-hand side 
should cancel out once we take advantage of the fact that in this limit $\sigma\ra 1$ 
(recall that $H=1/\ell$).

The conserved quantity associated with the Killing vector $\xi$ will be 
the mass and it can be evaluated at past or future infinity according to the sign of $H$. 
In our case the conserved mass is readily seen to be 
\beqs
\mathfrak{M}&=&-M,
\eeqs
which coincides, as expected with the value in a static coordinate system.
In particular, a divergent $\mathfrak{M}$ is found again for $M_H>M_{S}$ or 
 $M_W>M_b$. 

According to the dS/CFT conjecture, the dual field theory should live in our case on an 
Euclidian manifold whose metric is :
\beqs
ds^2&=&\gamma_{ij}dx^idx^j=dR^2+R^2d\Omega^2
\eeqs
obtained by an infinite conformal rescaling of the boundary geometry on $\partial{\cal M}$,
corresponding to a flat $D=3$ Euclidean metric. 
The stress energy tensor of the dual theory will be given by:
\beqs
8\pi G\tau^R_R&=&\frac{2M\ell}{R^3},~~~
8\pi G\tau^{\theta}_{\theta}=8\pi G\tau^{\varphi}_{\varphi}=-\frac{M\ell}{R^3}.
\eeqs
It is easy to see that this stress-tensor is covariantly conserved and traceless, 
as expected on general grounds.

\section{Conclusions}
This work was partially motivated by the question of how a positive
cosmological constant will affect the properties of a gravitating monopole.
To the best of our knowledge, this question has not yet been addressed in the literature,
except for the preliminary results in \cite{Brihaye:2005ft}.
 Apart from this motivation, the study of gravitating matter fields configurations
in asymptotically dS space may help a better understanding of the conjectured dS/CFT
correspondence.
 
We have found that despite the existence of a number of similarities to the $\Lambda=0$ case
the asymptotically dS solutions exhibits some new qualitative features.
All solutions present a cosmological horizon, on which matter fields take non-trivial values.
An interesting feature of the dS solutions appears to be
the absence of monopole configurations without a Higgs potential.
Also, contrary to the naive expectation
that a small $\Lambda$ will
not affect the properties of the configurations drastically,
we find that the mass of dS solutions evaluated at future/past infinity
by using the quasilocal
tensor of Brown and York diverges
(although the mass within the cosmological horizon stays finite).

Since this result is based only on the asymptotic expansion of the 
field equations together with the Higgs mechanism,
we expect it to remain valid for the dS versions of various
possible extensions of the magnetic monopole model.
Moreover, our results appear to be a generic feature of the asymptotically dS 
particle-like solutions of a  spontaneously
broken gauge theory.
Similar qualitative results are found for dS sphalerons with a Higgs doublet
\cite{Brihaye:2005ft}.

The question of the stability of our solutions is  
very important.
For the monopoles, we know that the solutions on the
first branch are stable for $\Lambda=0$, since
these are nothing else but gravitating monopoles 
in asymptotically flat space. Invoking
arguments from catastrophe theory \cite{stability}, 
we thus believe that the monopoles on the first branch
for $0 < \Lambda < \Lambda_{max}$ are stable, while 
the solutions on the second branch
have one mode of instability. 
This statement should definitely 
by checked by means of a detailed analysis, e.g. with the
one discussed in \cite{forgacs1}.
 
Supplementing the EYM system with a cosmological
constant leads to the occurrence of the "bag of gold" family
of solutions, where an equator naturally appears in the metric
and the spacetime becomes spatially compact and homeomorphic
to a three-sphere \cite{Volkov:1996qj}. Nothing like that was observed in the 
present
context because the Higgs function is imposed to approach its expectation
value asymptotically. Relaxing this condition, however, we think that it 
could  lead
to compact solutions as well for fine tuned values of the cosmological 
constant like e.g. in \cite{brihaye_delsate}. 
We plan to study this possibility for EYMH in near future.

Also, it would be interesting to apply the isolated horizon formalism, recently extended
to asymptotically dS spacetimes \cite{Corichi:2003kd}, to the particular system discussed here.

The divergencies described above  are not  too  disturbing for physics
inside the cosmological horizon.
However, a divergent mass-energy and action appear to   lead to 
severe problems   concerning the  possible holographic description \cite{Strominger:2001pn}
of a gravitating spontaneously broken gauge theory in dS spacetime.
 A divergent ADM mass has been found also for solutions of some theories
in asymptotically AdS spacetime.
However, in some cases it is still  possible to obtain a finite mass by
allowing the regularising counterterms to depend not only on
the boundary metric  but
also on the matter fields on the boundary \cite{Hollands:2005wt}.
It would be interesting to generalise this method to the dS case
and to assign a finite mass (evaluated outside the cosmological horizon)
to the solutions of a
spontaneously broken gauge theory.
We believe that this may lead to further understanding
of the rich structure of a field theory in dS space
as well as profound implications to the evolution of the
early universe.

It is likely that a re-examination of various
field theory nonperturbative effects for a dS ground 
state may lead to further surprises.
\\
\\
\\
\noindent
{\bf\large Acknowledgements} \\
We would like to thank Robert Mann for valuable remarks on a draft of this
paper.
YB is grateful to the
Belgian FNRS for financial support.
The work of ER is carried out
in the framework of Enterprise--Ireland Basic Science Research Project
SC/2003/390 of Enterprise-Ireland. 
The work of C.S. was supported by the Natural Sciences and Engineering Research Council of Canada.

\vspace*{1.cm}
\textbf{\Large Appendix A: BPS solutions with a Liouville potential}
\\
\\
In \cite{Gibbons:1993xt}
it was shown that the   BPS monopoles of four-dimensional YMH 
theory continue to be exact solutions of this model 
even after the inclusion of gravitational, electromagnetic and 
dilatonic interactions, 
provided that certain non-minimal interactions are included.
The solutions presented in \cite{Gibbons:1993xt} are asymptotically flat.
Here we prove that a similar construction can be done for a 
Liouville dilaton potential, $i.e.$ a positive {\it effective}
cosmological constant.

Following the  conventions
in \cite{Gibbons:1993xt},
we consider a four dimensional action principle on the form
\begin{eqnarray}
\label{S0}
S &=& \frac{1}{4\pi G} \int d^4 x \sqrt{-g}  
\bigg[ \frac{R}{4}-\frac{1}{2}(\nabla \sigma)^2-e^{2b \sigma}\frac{1}{4}f_{\mu \nu}f^{\mu \nu}
-\frac{ e^{-2b \sigma}}{4}\Lambda\bigg]+S_{matter}
\end{eqnarray}
with $\sigma$ the dilaton field\footnote{The dilaton
$\sigma$ should not be confused with
the metric function $\sigma(r)$ which enters the ansatz (\ref{metric}).}
$b$ the dilaton coupling constant and $f_{\mu \nu}$ the U(1) field.
Different from the situation in \cite{Gibbons:1993xt}, the above action
principle contains a Liouville potential $V(\sigma)= -e^{-2b \sigma} \Lambda/4$.
The expression of the matter action is given by 
\begin{eqnarray}
\label{Sm}
S_{matter} =   \int\!\! d^4x
\sqrt{-g}
\bigg\{ -{1\over 2} e^{{(1-b^2)\over b}\sigma}  
{\rm Tr}(F_{\mu\nu}F^{\mu\nu}) -{1\over 4} e^{-{(1+b^2)\over b}\sigma}  
{\rm Tr}(D_\mu\Phi D^\mu\Phi)   + \frac{c\sqrt{1+b^2}}{4\sqrt{-g}}
\varepsilon^{\mu\nu\alpha\beta}f_{\mu\nu} (\Phi
F_{\alpha\beta}) \bigg\}   ,
\end{eqnarray}
with $F_{\mu\nu}$ the SU(2) field, $\Phi$ the Higgs scalar, while
 $c$ is a constant defined bellow.
 Also, to simplify the general picture,
  we take here the gauge coupling constant $g=1$.
 
The Einstein-Maxwell-dilaton field equations are given by
\begin{eqnarray}
 G_{\mu\nu}
-2T_{\mu\nu}(f) -2T_{\mu\nu}(\sigma) &= (8\pi G)T_{\mu\nu}(mat.)
\cr
\nabla_\mu\big( e^{2b\sigma}f^{\mu\nu}\big) &= (4\pi G) J^\nu(mat.)
\cr
\partial_\mu\big( \sqrt{-g}g^{\mu\nu}\partial_\nu\sigma\big)
-{g\over2}\sqrt{-g}e^{2b\sigma} f^{\mu\nu}f_{\mu\nu} &= -(4\pi G){\delta
S_{matter}\over\delta\sigma} 
\end{eqnarray}
where we note
\begin{eqnarray}
\nonumber
&&
T_{\mu\nu}(f)  = e^{2b\sigma}\left( f_{\mu\lambda}f_\nu{}^\lambda
-{1\over4} g_{\mu\nu}f^2\right),
~~~T_{\mu\nu}(\sigma) =
\left(\partial_\mu\sigma\partial_\nu\sigma
-{1\over2}g_{\mu\nu}(\partial\sigma)^2-g_{\mu \nu}V(\sigma)\right), 
\\
&& T_{\mu\nu}(mat.) = e^{{(1-b^2)\over b}\sigma} {\rm Tr}
\left( F_{\mu\lambda}F_{\nu}{}^{\lambda}-{1\over 4}g_{\mu\nu}
F_{\alpha\beta}F^{\alpha\beta}\right) 
+ e^{-{(1+b^2)\over b}\sigma} {\rm Tr}\left( D_{\mu}\Phi D_{\nu}\Phi
-{1\over 2}g_{\mu\nu}D_{\alpha}\Phi D^{\alpha}\Phi \right)\cr
&&J^{\nu}(mat.)=
{ce^{-1}\over 2 }
\sqrt{1+b^2}\varepsilon^{\nu\mu\rho\sigma}\tr\left( D_{\mu}\Phi
G_{\rho\sigma}\right)\ .    
\end{eqnarray}
In \cite{Gibbons:1993xt} it has been shown that
for $\Lambda=0$
the equations of motion for $F_{\mu\nu}$ and $\Phi$ are solved by any 
YMH configuration that solves the   {\it flat space} Bogomol'nyi equations, 
\begin{eqnarray}
G_{ij} =\mp  ~\varepsilon^{ijk}D_k\Phi\ , 
\end{eqnarray}
provided $c^2=1$.
To this aim, a special metric ansatz has been used, in terms of only one 
metric function.
The the Einstein, Maxwell and dilaton equations are
then equivalent to a single Euclidean
3-space Poisson equation.
In the absence of matter fields $S_{matter}=0$,
this choice gives the "extreme" dilaton black hole solutions of \cite{Gibbons:1982ih}.

This construction can be generalised for a Liouville potential term in the action principle.
However, given the presence of an effective cosmological constant, the
metric ansatz would be time-dependent.

A straightforward generalisation of the metric ansatz used in  \cite{Gibbons:1993xt} is 
\begin{eqnarray}
\label{s0}
ds^2=-e^{-2\phi}dt^2+R^2(t)e^{2\phi}d{\bf x}\cdot d{\bf x},
~~~{\rm where}~~~
e^{\phi}=e^{\sigma/b}/R(t),~~~ 
R(t)=(\frac{t}{t_0})^{1/b^2},
\end{eqnarray}
the only nonvanishing component of the $U(1)$ potential being
\begin{eqnarray}
\label{A0}
A_t=\frac{1}{\sqrt{1+b^2}}R(t)^{-b^2}e^{-(1+b^2)\phi}.
\end{eqnarray}
 The free parameter $t_0$ is fixed by $\Lambda$
\begin{eqnarray}
\label{t0}
\Lambda=\frac{2}{b^2}(\frac{3}{b^2}-1)\frac{1}{t_0^2}.
\end{eqnarray}
It is convenient to define
\begin{eqnarray}
\label{t0_2}
e^{\phi}=(1+R^{-(1+b^2)}f(\bf x))^{1/(1+b^2)}.
\end{eqnarray}
A direct computation shows that, for $c^2=1$, the YMH equations
implied by the action principle (\ref{Sm})   
are automatically satisfied  by any flat space solution of the
Bogomol'nyi equations.
 
Then, the function $f(\bf x)$ is uniquely determined as 
a solution of the equation
\begin{eqnarray}
\label{wq0}
\nabla^2f=-4\pi G(1+b^2) \tr\left( D_i\Phi\cdot
D_i\Phi\right),
\end{eqnarray}
which follows from the Einstein-Maxwell-dilaton equations.
In the absence of $S_{matter}$, one recovers the  
(multi-)black hole solutions in cosmological Einstein-Maxwell-dilaton 
theory discussed in \cite{Maki:1992tq}.


\newpage

\newpage

\newpage

\newpage

\newpage

\newpage
\newpage
\begin{figure}
\epsfysize=12cm
\epsffile{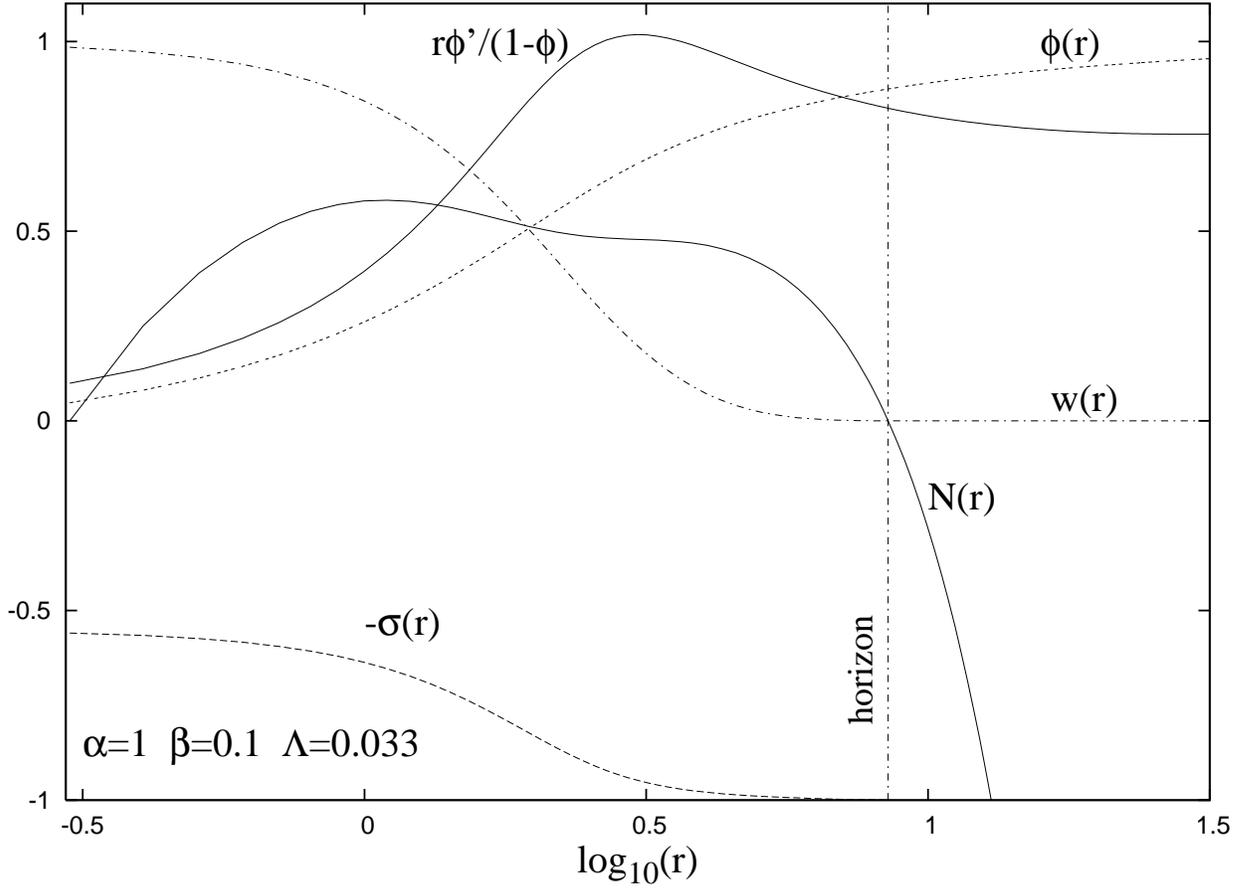}
\vskip 3cm
\caption{\label{eym8}
 A typical black hole solution in EYMH theory is plotted for 
$\alpha=1$,~$r_h=0.3$
and $\Lambda=0.033$.
We also plot $r\phi'/(1-\phi')$ which indicates the power with which
$\phi$ reaches its asymptotic value.
}
\end{figure}

\newpage
\begin{figure}
\epsfysize=12cm
\epsffile{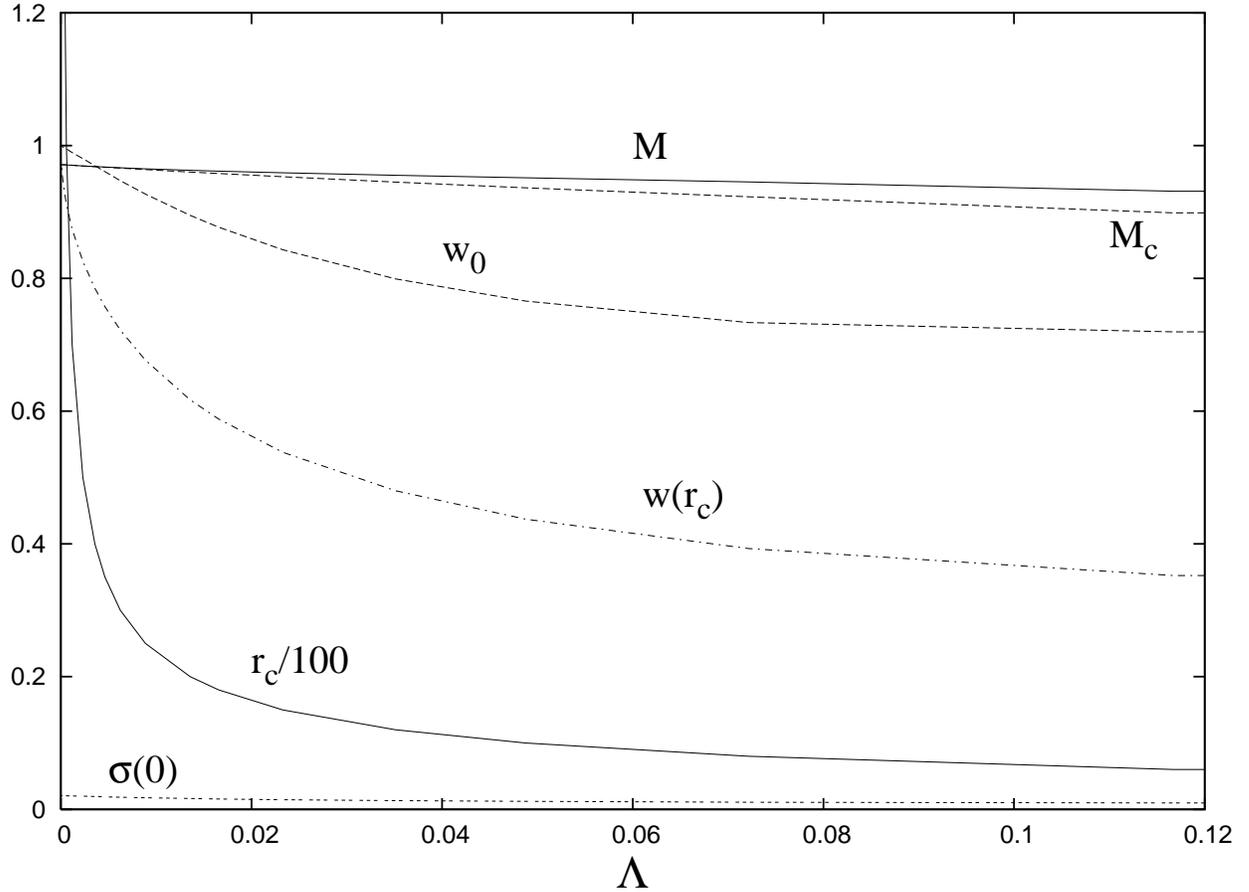}
\vskip 3cm
\caption{\label{eym-spectrum} 
Several characteristic features of the two node particle-like solutions
of the EYM-$\Lambda$ system are plotted as a function of the cosmological
constant.
Here $M$ and $M_c$ are the values of the mass function $m(r)$ at infinity and at the
cosmological horizon, $r_c$ is the event horizon radius and $\sigma(0)$ is the value 
of the metric function $\sigma$ at the origin.
Also,  $w_0$ and $w(r_c)$ are the values of the magnetic gauge potential at infinity and at the
cosmological horizon.
}
\end{figure}

\newpage
\begin{figure}
\epsfysize=12cm
\epsffile{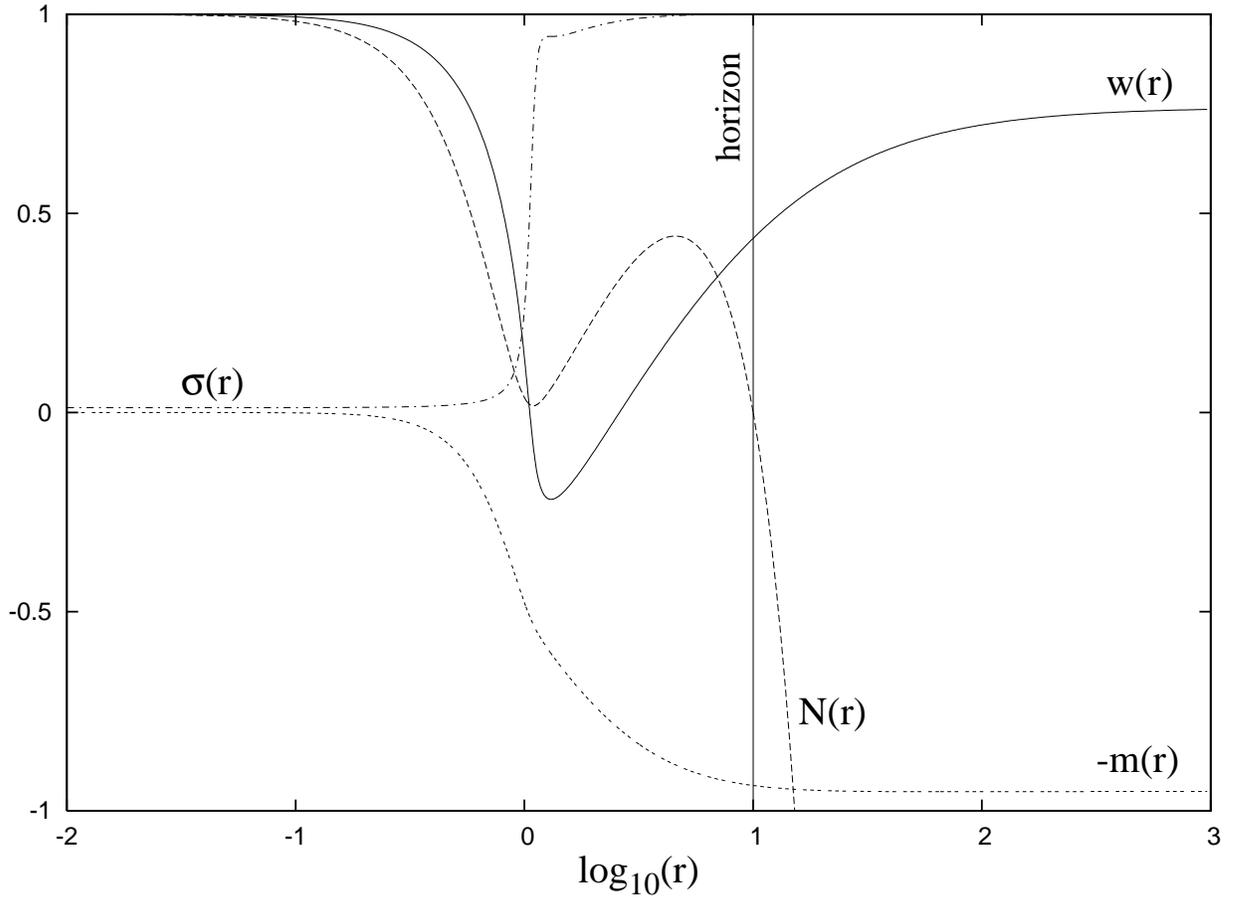}
\vskip 3cm
\caption{\label{eym-2nodes} 
A typical asymptotically de Sitter monopole 
solution with a regular origin in EYM-$\Lambda$ theory.
The value at infinity of the magnetic gauge potential is $w_0\simeq 0.761$.
}
\end{figure}

\newpage
\begin{figure}
\epsfysize=12cm
\epsffile{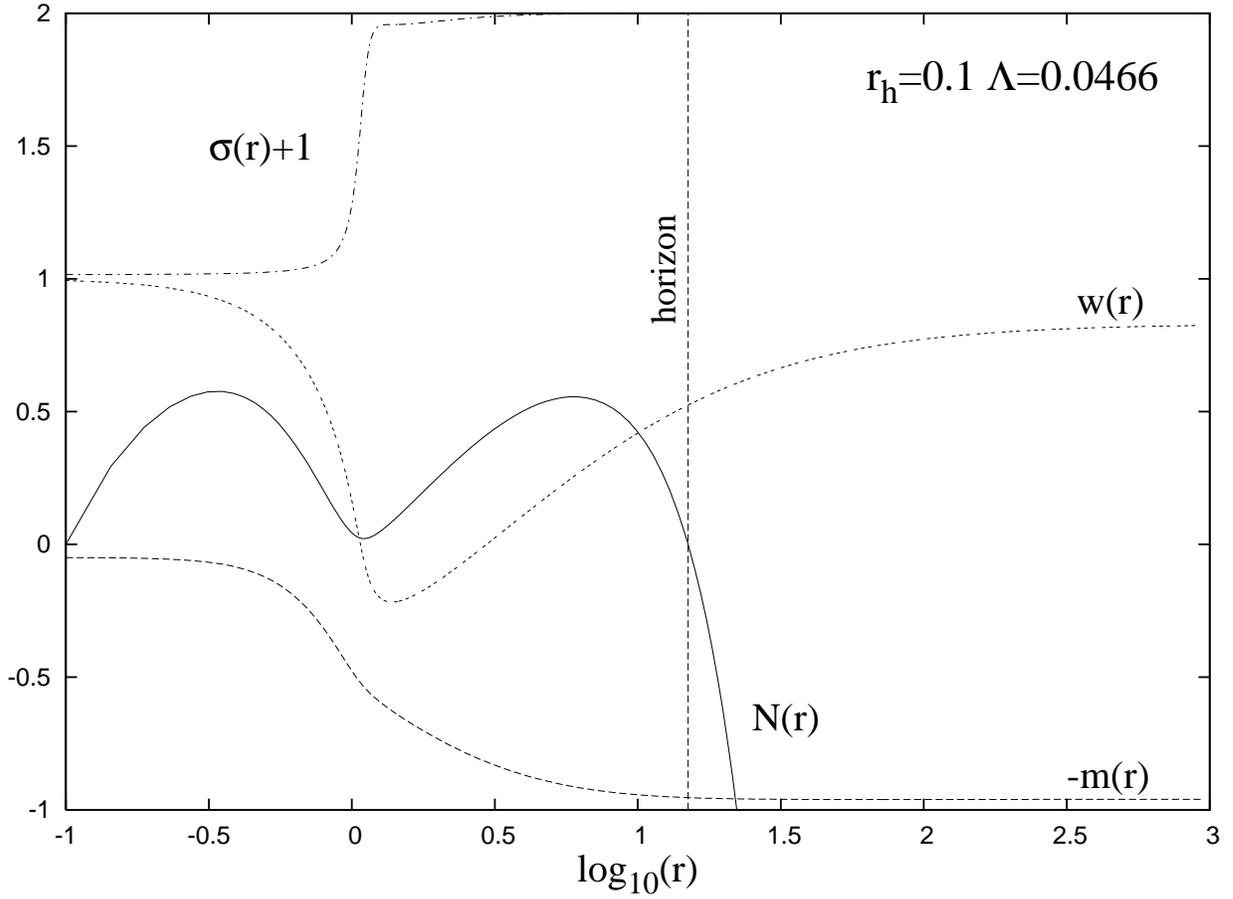}
\vskip 3cm
\caption{\label{eym-bh} 
A typical asymptotically de Sitter black hole monopole solution in EYM-$\Lambda$ theory.
The cosmological horizon radius is $r_c=15$.
The value of the gauge field
on the event horizon is $w_h=0.9936$, while at infinity $w_0=0.8244$.
}
\end{figure}

\end{document}